\documentclass[twocolumn,showpacs,preprintnumbers,amsmath,amssymb]{revtex4}
\usepackage{graphicx}
\graphicspath{{./figs/}}
\usepackage{dcolumn}
\usepackage{bm}
\usepackage{float}
\usepackage[Gray,squaren,thinqspace,thinspace]{SIunits} 

\usepackage{grffile}
\newcommand{\ket}[1]{\mid #1 \rangle}

\newcommand{\braket}[2]{\langle #1 \mid #2 \rangle}
\begin{document}

\title{Quantifying short-range correlations in nuclei}
\author{Maarten Vanhalst}
\email{Maarten.Vanhalst@UGent.be}
\author{Jan Ryckebusch}
\email{Jan.Ryckebusch@UGent.be}
\author{Wim Cosyn}
\email{Wim.Cosyn@UGent.be}

\affiliation{Department of Physics and Astronomy,\\
 Ghent University, Proeftuinstraat 86, B-9000 Gent, Belgium}
\date{\today}
\begin{abstract}

\begin{description}
\item[Background] Short-range correlations (SRC) are an important ingredient of
  the dynamics of nuclei. 
\item[Purpose] An approximate method to quantify the magnitude of the
  two-nucleon (2N) and three-nucleon (3N) short-range correlations
  (SRC) and their mass dependence is proposed.
\item[Method] 
  The proposed method relies on the concept of the ``universality'' or
  ``local nuclear character'' of the SRC. We quantify the SRC by computing the
  number of independent-particle model (IPM) nucleon pairs and triples
  which reveal beyond-mean-field behavior. It is argued that those can
  be identified by counting the number of nucleon pairs and triples in a zero
  relative orbital momentum state. A method to determine the quantum
  numbers of pairs and triples in an arbitrary mean-field basis is
  outlined.
\item[Results] 
  The mass dependence of the 2N and 3N SRC is studied.  The
  predictions are compared to measurements. This includes the ratio of the inclusive inelastic
  electron scattering cross sections of nuclei to $^{2}$H and $^{3}$He
  at large values of the Bjorken variable. Corrections stemming from the center-of-mass motion of the pairs are estimated. 
\item[Conclusions] We find that the relative probability per nucleon
  for 2N and 3N SRC has a soft dependence with mass number $A$ and
  that the proton-neutron 2N SRC outnumber the proton-proton
  (neutron-neutron) 2N SRC. A linear relationship between the
  magnitude of the EMC effect and the predicted number of
  proton-neutron SRC pairs is observed. This provides support for the
  role of local nuclear dynamics on the EMC effect.
\end{description}
\end{abstract}
%
\pacs{25.30.Fj,24.10.-i,13.60.Hb}

\maketitle 
\section{Introduction}
We define the nuclear packing factor (NPF) as the fraction of the
nuclear volume that is occupied by nucleons.  A rough order of
magnitude estimate of the NPF can be arrived at using uniform spheres
for the nuclear and nucleon density.  The nuclear radius $ R_{A} $ can
be reasonably determined from $R_{A} = 1.2$(fm)~$ A ^{ 1/3}$. It is
not obvious what value of the nucleon radius $r_N$ should be used. In
models of relativistic heavy-ion collisions it is customary
\cite{ISI:000279267600001} to use expulsion distances $d$, which
simulate the hard-core NN repulsion, of the order of 1~fm,
corresponding with $r_N \approx $0.5~fm. This leads to NPF=0.07.  A
recent reanalysis of electron scattering data resulted in a
root-mean-square charge radius of the proton $r_{p}^{c} = \sqrt{
  \left< r_p ^2\right>}$=0.897(18)~fm \cite{PhysRevC.72.057601}.
Assuming that the $r_{p}^{c}$ is an estimate of the proton and neutron
radius one arrives at NPF= $\left( \frac{r_p^{c} \textrm{(fm)}} {1.2}
\right)^{3}$=0.42.  It is clear that the computed NPF is very
sensitive to the adopted value of the nucleon radius. The estimate of
the NPF on the basis of $r_{p}^{c}$ should be considered as an upper
limit. Indeed, the established value of the nuclear saturation density
of 0.17~nucleons/fm$^{³}$ corresponds with a mean internucleon
distance of 1.8~fm implying that $r_{N} \le 0.9$~fm.

From the above, it is clear that one expects that the nucleus is
more like a saturated quantum liquid than a gas of freely moving
nucleons.  Accordingly, the nuclear wave functions receive large
corrections from short-range (SRC) and long-range correlations. These days
it is common practice to implement the effect of SRC in nuclear
computations. Examples include the calculations of matrix elements for
double-$\beta$ decay \cite{Engel:2011ss}, of event simulations in
heavy-ion collisions \cite{Alvioli:2010yk}, and of hadron
transparencies in nuclei \cite{Cosyn:2007er}.

The EMC effect \cite{Aubert:1983xm} is the reduction of the cross
section for leptonic scattering off a nucleon bound in a nucleus
relative to that of a free nucleon (mass $M_N$). The EMC effect was
observed in Deep Inelastic Scattering (DIS) experiments on nuclei at
high virtual-photon virtualities $Q^{2} = q ^{2} - \omega ^{2} \gtrsim
2$~GeV$^{2}$ for Bjorken $x_{B} = \frac {Q ^ 2} { 2 M_N \omega} $ in
the range $0.3 \leq x_{B} \leq 0.7$. The ratio of per nucleon cross
sections is denoted by $ R = \frac {2} {A} \frac {\sigma ^{A} }
{\sigma ^{D}}$ where $\sigma ^{A}$ is the cross section for leptonic
scattering from the target $A$. The magnitude of the EMC effect can be
quantified by means of the slope $ - \frac {d R} {dx _{B} }$
\cite{Seely:2009gt}. Another remarkable feature of the ratio $R$ is
that it adopts a constant value (this factor is commonly referred to
as the SRC scaling factor $ a_{2} (A /D) $) for $1.5 \lesssim x_{B}
\lesssim 2$ and moderate values of $Q^{2}$ \cite{Day:1987az,
  PhysRevLett.96.082501, PhysRevLett.108.092502}.  It has been
suggested \cite{Frankfurt:2008zv} that the $ a_{2} (A /D) $ can be
related to the high-momentum components of the nuclear wave functions.
A phenomenological linear relationship between the $ a_{2} (A /D) $
and the magnitude of the EMC effect expressed as $ - \frac {d R} {dx
  _{B} }$ has been observed
\cite{Weinstein:2010rt,Hen:2012fm,Arrington:2012ax}.  This indicates
that the magnitude of the European Muon Collaboration (EMC) effect may
be driven by SRC.  In this picture the magnitude of the EMC effect is
(partly) related to the temporal local density fluctuations which are
induced by the high virtualities of the leptonic probe.  Recent
measurements \cite{Seely:2009gt} corroborate this relation between the
local nuclear environment and the magnitude of the EMC effect.

Given an arbitrary nucleus $A(N,Z)$ we address the issue of
quantifying the number of two-nucleon (2N) pairs prone to SRC and the
number of 3N triples prone to SRC.  Along the same lines we
investigate to what extent the mass dependence of the NN SRC
can be captured by some approximate principles.  We wish to develop a
robust method which is applicable to any nucleus from He to Pb.  From
this method we expect, for example, that it allows one to study the
mass dependence of the SRC without combining results from various
types of calculations.

Momentum distributions contain the information about 1N, 2N, 3N,
$\ldots$ properties of the nuclear ground state. Over the years
various methods to compute the nuclear 1N and 2N momentum
distributions have been developed. Ab-initio calculations which solve
the Schr\"{o}dinger equation with realistic nucleon-nucleon
interactions are available for light nuclei like $^{4}$He
\cite{Schiavilla:2006xx,Wiringa:2008dn,Feldmeier:2011qy}. For
medium-weight nuclei (12 $\le $ A $\le$ 40) truncation schemes based
on cluster expansions can be adopted \cite{Alvioli:2007zz}.
Correlated-basis function theory has been applied to compute
ground-state densities and momentum distributions for
doubly-closed-shell nuclei from $^{12}$C to $^{208}$Pb
\cite{Bisconti:2006hv, Bisconti:2007vu}.  Thanks to the enormous
progress in theoretical many-body nuclear physics and the availability
of nuclear momentum distributions in a broad mass range, times are
ripe to learn more about SRC, for example by mapping its $A$ and
isospin dependence.  It remains notoriously difficult, though, to
establish quantitative relationships between observables and the
computed momentum distributions
\cite{Frankfurt:2008zv,Vanhalst:2011es,Arrington:2011xs,
  shneor:072501,jan2002,Ryckebusch:1996wc}.  Here, we do not attempt a
high-precision calculation of momentum distributions. Our goal is to
gather insight into the mass and isospin dependence of the SRC from
stylized facts of momentum distributions.

In a mean-field model fluctuations are completely ignored.  The SRC
induce spatio-temporal fluctuations from the mean-field predictions
for the nuclear density distributions for example. As a result of SRC,
realistic nuclear wave functions reflect the coexistence of single
nucleon (mean-field) structures and cluster structures. The clusters
account for beyond mean-field behavior. As the nucleon-nucleon
interaction is short ranged, the clusters attributed to SRC are
predominantly 2N.  The central result of this paper asserts that the
amount of 2N and 3N SRC in nuclei can be reasonably quantified by
counting the number of nucleon pairs and triples in a zero relative
orbital state in a mean-field ground-state wave function. In order to
quantify the isospin dependence of the 2N and 3N correlations,
additional information about the spin dependence of the clusters is
necessary.

This paper is organized as follows. Sect.~\ref{sec:II} is devoted to a
discussion of momentum distributions and of how they
can be used to quantify the mass and isospin dependence of SRC. In
Sect.~\ref{sec:results} we address the issue whether inclusive
electron scattering data can be linked to the number of correlated 2N
and 3N clusters. Thereby, we deal with both the $a_{2}(A/D)$ coefficient 
and the magnitude of the EMC effect.

\section{Quantifying nuclear correlations}
\label{sec:II}
In this section we start from stylized facts of nuclear momentum
distributions in order to arrive at criteria to quantify the 2N and 3N
SRC in nuclei.  Our focus is on their mass dependence.

\subsection{Nuclear momentum distributions}
\label{sec:IIA}
In this subsection we provide the definitions and normalization
conventions of the nuclear momentum distributions used here.  For the
sake of the simplicity of the notations, we will only consider the
positional degrees-of-freedom. Unless stated otherwise the spin - and
isospin degrees-of-freedom are not explicitly written in the
expressions.

The one-body momentum distribution of nuclei is defined as 
\begin{equation}
P_1 \left( \vec{k} \right) = 
\frac {1} { \left( 2  \pi \right) ^{3} }   
\int d \vec{r}_{1} 
\int d \vec{r}_{1} ^{\; \prime} 
e ^{i \vec{k} \cdot \left( \vec{r}_{1} - \vec{r}_{1}^{\; \prime} \right) } 
\rho _{1} \left(   
\vec{r}_{1} , \vec{r}_{1} ^{\; \prime} \right) \; ,
\label{eq:momonebody}
\end{equation}
where $ \rho _{1} \left( \vec{r}_{1} , \vec{r}_{1} ^{\; \prime}
\right) $ is the one-body non-diagonal density matrix
\begin{eqnarray}
  \rho _{1} \left( \vec{r}_{1} , \vec{r}_{1} ^{\; \prime} \right) & = & 
     \int \left\{ d \vec{r} _{2 -N} \right\} 
  \Psi ^{*} _{A}  \left(\vec{r}_1, \vec{r} _2, \vec{r}_3, \ldots, \vec{r}_A   \right) \nonumber \\
& & \times
  \Psi _{A} \left(\vec{r}_1 ^{\; \prime}, \vec{r} _2 , \vec{r}_3, \ldots, \vec{r}_A   \right)  \; .
\label{eq:mom1b}
\end{eqnarray}
Here, $ \Psi _{A} $ is the ground-state wave function of the nucleus
$A$ and the notation
\begin{equation}
\left\{ d \vec{r} _{i -N} \right\} = d \vec{r}_i d \vec{r}_{i+1} \ldots d \vec{r}_A \; ,
\end{equation}
has been introduced.
For $\left< \Psi _{A} \right| \left. \Psi _{A}
\right> = 1 $, one has that
\begin{equation}
\int d \vec{k}  P_1 \left( \vec{k} \right) = 1 \; .
\label{eq:normap1k}
\end{equation}

We introduce relative and center-of-mass (c.m.) coordinates of nucleon pairs
in coordinate $\left(\vec{r}_{12}, \vec{R}_{12} \right)$ and momentum
space $\left(\vec{k}_{12}, \vec{P}_{12} \right)$
\begin{equation}
\vec{r}_{12} = \frac {\vec{r} _1 - \vec{r} _2 } {\sqrt{2} }
 \hspace{0.1\textwidth}
\vec{R}_{12} = \frac {\vec{r} _{1} + \vec{r} _{2} } { \sqrt{2}}
\label{eq:cor1}
\end{equation}
\begin{equation}
\vec{k}_{12} = \frac{ \vec{k} _1 - \vec{k} _2 } {\sqrt{2}} \hspace{0.1\textwidth}
\vec{P}_{12} = \frac {\vec{k} _{1} + \vec{k} _{2} } {\sqrt{2}} \; , 
\label{eq:cor2}
\end{equation}
and define the two-body momentum distribution in the standard fashion as
\begin{eqnarray}
P_2 \left( \vec{k}_{12}, \vec{P}_{12} \right) & = & 
\frac {1} { \left( 2  \pi \right) ^{6} }   
\int d \vec{r}_{12} \int d \vec{R}_{12}
\int d \vec{r}_{12} ^{\; \prime} \int d \vec{R}_{12} ^{\; \prime}
\nonumber \\
& & \times 
e ^{i \vec{k}_{12} \cdot \left( \vec{r}_{12} - \vec{r}_{12}^{\; \prime} \right) } 
e ^{i \vec{P}_{12} \cdot \left( \vec{R}_{12} - \vec{R}_{12}^{\; \prime} \right) } 
\nonumber \\
& & \times 
\rho _{2} \left(   
\vec{r}_{12} , \vec{R}_{12} ;
\vec{r}_{12} ^{\; \prime} , \vec{R}_{12} ^{\; \prime}
\right) \; .
\label{eq:mom1}
\end{eqnarray}
Here, $ \rho _{2} \left( \vec{r}_{12} , \vec{R}_{12} ; \vec{r}_{12} ^{\; \prime} ,
  \vec{R} _{12} ^{\; \prime} \right) $ is the two-body non-diagonal density matrix
\begin{widetext}
\begin{eqnarray}
\rho _{2}  \left( \vec{r}_{12} , \vec{R}_{12} ; \vec{r}_{12} ^{\; \prime},
    \vec{R} _{12} ^{\; \prime} \right) 
& = &  
  \rho _{2} \left( 
    \vec{r}_1 = \frac {+ \vec{r}_{12}  + \vec{R} _{12}} {\sqrt{2}}  ,
    \vec{r}_2 = \frac {- \vec{r}_{12}  + \vec{R} _{12}} {\sqrt{2}}  ; 
    \vec{r}_1 ^{\; \prime} = \frac {+ \vec{r}_{12} ^{\; \prime} 
                                 + \vec{R}_{12} ^{\; \prime}} {\sqrt{2}} ,
    \vec{r}_2 ^{\; \prime} = \frac {- \vec{r}_{12} ^{\; \prime} 
                                 + \vec{R}_{12} ^{\; \prime}} {\sqrt{2}}
  \right) 
  \nonumber \\
  & = & \int \left\{ d \vec{r} _{3 -N} \right\} 
  \Psi ^{*} _{A}  \left(\vec{r}_1, \vec{r} _2, \vec{r}_3, \ldots, \vec{r}_A   \right) 
  \Psi _{A} \left(\vec{r}_1 ^{\; \prime}, \vec{r} _2 ^{\; \prime}, \vec{r}_3, \ldots, \vec{r}_A   \right)  \; .
\label{eq:twobodydensity}
\end{eqnarray}
\end{widetext}
One has the normalization condition 
\begin{equation}
\int d \vec{k}_{12} \int d \vec{P}_{12} P_2 \left( \vec{k}_{12}, \vec{P}_{12} \right) = 1 \; .
\end{equation}
In a spherically symmetric system, the two-body momentum distribution
$ P_2 \left( \vec{k}_{12}, \vec{P}_{12} \right) $ depends on three
independent variables. One of the most obvious choices
\cite{Alvioli:2011aa} is
\begin{equation} 
\left( 
\mid \vec{k}_{12} \mid, \mid \vec{P}_{12} \mid , \theta _ {\vec{k}_{12} \vec{P}_{12} }
\right) \; ,
\end{equation}
where $ \theta _ {\vec{k}_{12} \vec{P}_{12} }$ is the angle between
$\vec{P}_{12}$ and $\vec{k}_{12}$.

The distributions $P_1 \left( \vec{k} \right) $ and $ P_2 \left(
  \vec{k}_{12}, \vec{P}_{12} \right) $ reflect all information about
one-nucleon and two-nucleon properties contained in the ground-state
wave function. Other quantities can be directly related to them. Here,
we list some of the most frequently used ones. 

The two-body c.m. momentum distribution is defined as $ \left( d
  \vec{P}_{12} = P_{12} ^2 d P_{12} d \Omega _{{P}_{12}} \right)$
\begin{equation}
P_{2} (P_{12}) = \int d \vec{k}_{12} \int d \Omega _{P_{12}} P_2 \left( \vec{k}_{12}, \vec{P}_{12} \right) \; .   
\label{eq:comdis}
\end{equation}  
The quantity $P_{12} ^{2} P_{2} \left(P_{12} \right) dP_{12} $ is related to the
probability of finding a nucleon pair in $A$ with c.m. momentum $P_{12} =
\mid \vec{P}_{12} \mid $ irrespective of the value and direction of the relative momentum
$ \vec{k}_{12} $ of the pair. The $P_{2} \left( P_{12} \right)$ receives contributions from the
proton-proton, neutron-neutron, and proton-neutron pairs
\begin{equation}
P_{2} \left(P_{12}\right) = P_{2} ^{pp} \left(P_{12} \right) 
                        + P_{2} ^{nn} \left(P_{12} \right) 
                        + P_{2} ^{pn} \left(P_{12} \right) \; .
\end{equation}  

In a spherically symmetric nucleus, it is convenient to introduce the
quantities
\begin{eqnarray}
  n_1 \left( k \right) & = & \int d \Omega _{k} P_1 \left( \vec{k} \right) \; , 
\label{eq:n1k}
\\
  n_2 \left( k_{12}, P_{12} \right) & = & 
  \int d \Omega _{k_{12}}   \int d \Omega _{P_{12}}   
  P_2 \left( \vec{k}_{12}, \vec{P}_{12} \right) \; .
\label{eq:n2k}
\end{eqnarray}
The quantity $n_1 \left( k \right) k ^{2} dk$ gives the probability of
finding a nucleon with a momentum in the interval $[k, k + d k]$. The
$ n_2 \left( k_{12}, P_{12} \right) k _{12}^{2} dk_{12} P _{12}^{2}
dP_{12}$ is the combined probability of finding a nucleon pair with a
relative momentum in $\left[ k_{12}, k_{12}+ d k_{12} \right]$ and
c.m. momentum in $\left[ P_{12}, P_{12}+ d P_{12} \right]$.

\subsection{Mean-field approximation and beyond}
\label{sec:IIB}
A time-honored method to account for the effect of correlations in classical
  and quantum systems is the introduction of correlation functions.
  Realistic nuclear wave functions $\mid { \Psi} \rangle $ can be
  computed after applying a many-body correlation operator to a Slater
  determinant $\mid \Psi ^{MF}  \rangle$
\begin{equation}   
 \mid { \Psi_A}   \rangle =  \frac{1}
{ \sqrt{\langle \ \Psi  ^{MF} _A \mid \widehat{\cal
G}^{\dagger} \widehat{\cal G} \mid \Psi  ^{MF} _A \ \rangle}} \ 
\widehat
{ {\cal G}} \mid  \Psi  ^{MF} _A \ \rangle \; .
\label{eq:realwf}
\end{equation}
The nuclear correlation operator $\widehat{\cal G}$ is complicated but
as far as the short-range correlations are concerned, it is dominated
by the central, tensor and spin correlations \cite{Roth:2010bm}
\begin{eqnarray}
\widehat{\mathcal{G}}  & \approx &  \widehat {{\cal S}}  
\biggl[ \prod _{i<j=1} ^{A} \biggl(
1 - g_c(r_{ij}) + f_{t\tau}(r_{ij}){S_{ij}} \vec{\tau}_i \cdot \vec{\tau}_j \nonumber \\
& + & 
f_{s\tau}(r_{ij}) \vec{\sigma}_i \cdot \vec{\sigma}_j \; \;  \vec{\tau}_i \cdot \vec{\tau}_j 
\biggr) \biggr] \; , 
\end{eqnarray}
where $g_c(r_{12})$, $f_{t\tau}(r_{12})$, $f_{s\tau}(r_{12})$ are the
central, tensor, and spin-isospin correlation function, $ \widehat
{{\cal S}} $ the symmetrization operator and ${S_{12}}$ the tensor
operator
\begin{eqnarray}
{S_{12}} & = & \frac {3} {r_{12}^2} \vec{\sigma} _{1} \cdot {\vec{r}_{12}}
\; \;               \vec{\sigma} _{2} \cdot {\vec{r}_{12}}
                  - \vec{\sigma} _{1} \cdot \vec{\sigma} _{2} 
=  \sqrt { \frac { 24 \pi} {5} } \sum _ { M_{L} } \left( -1 \right)^{M_{L} }
\nonumber \\
& & \times  Y_{2 M_{L} } \left( \Omega _{r_{12}} \right) 
\left[   \vec{\sigma} _{1}  \otimes \vec{\sigma} _{2}  \right] _{2 - M _{L} } 
\; . 
\end{eqnarray}
The operator $ {S_{12}} $ admixes relative two-nucleon states of
different orbital angular momentum, is operative on triplet spin
states only, and conserves the total angular momentum of the pair.

We stress that the correlation functions cannot be considered as
universal and that in some many-body approaches, particularly for
light nuclei, they do not appear. The momentum distributions which
result from the calculations depend on the interplay between many
factors, including the choices made with regard to the nucleon-nucleon
interaction, the single-particle basis (if applicable), the many-body
approximation scheme, $\ldots$. As a matter of fact, different
nucleon-nucleon interactions and many-body approaches may produce,
particularly in the region of SRC (short distances/high momenta),
momentum distributions which are very similar (see, e.g. Refs.
\cite{Schiavilla:2006xx, Feldmeier:2011qy, Pieper:1992gr,
  AriasdeSaavedra:1996kd}).

\begin{figure*}
\includegraphics[width=0.7\textwidth,angle=-90]{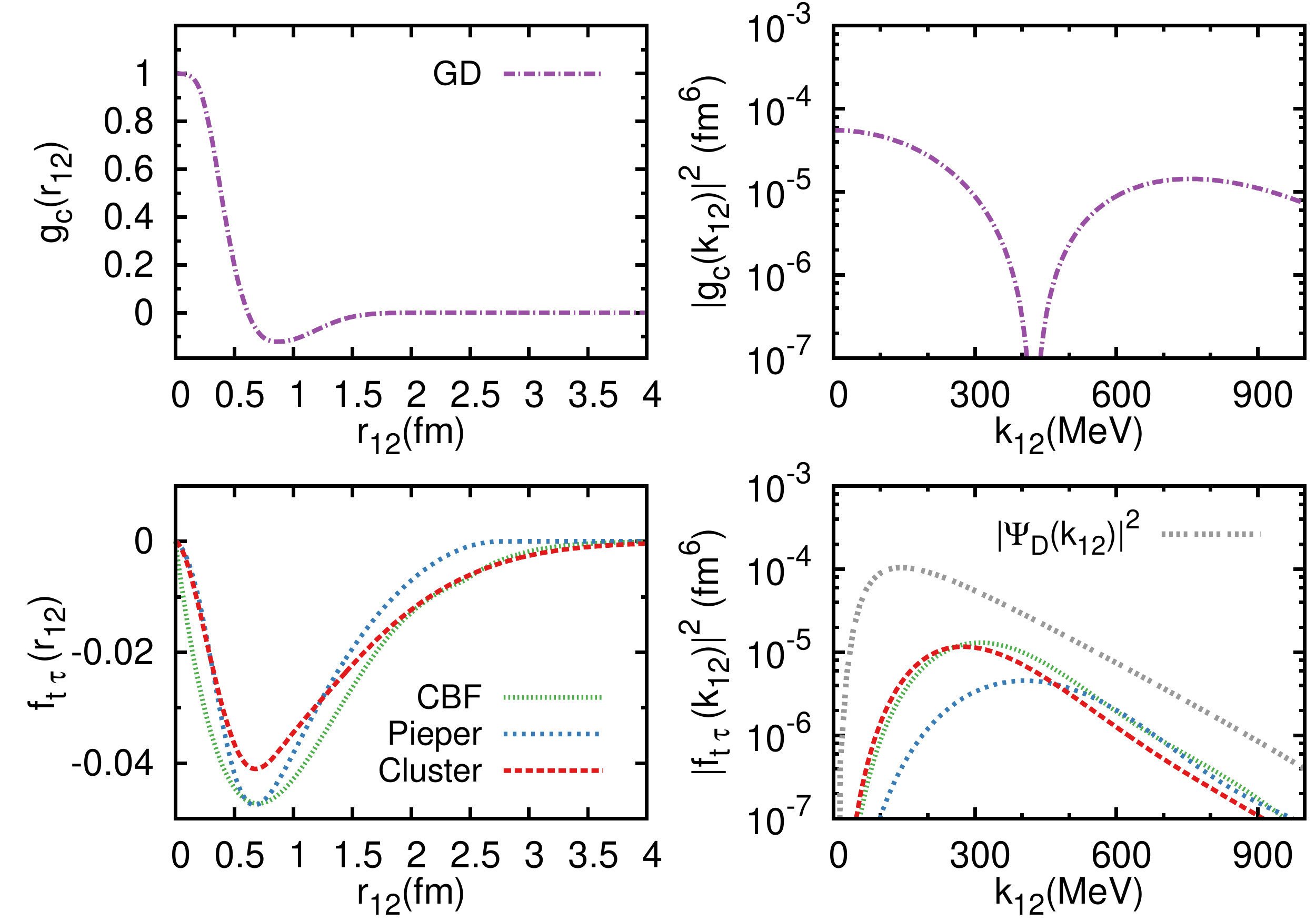}
\caption{(color online). The radial and momentum dependence of a
  central and some tensor correlation functions. 
  The central correlation function ``GD'' is for nuclear matter and from
  Ref.~\cite{gearheart94}.
  The tensor correlation function ``Pieper'' is for $^{16}$O and from Ref.~\cite{Pieper:1992gr}, the ``CBF'' one is for $^{16}$O and from
  Ref.~\cite{AriasdeSaavedra:1996kd}, and the ``cluster'' one is for
  $^{16}$O and from Ref.~\cite{Alvioli:2005cz}, $\Psi_D \left(k_{12}
  \right)$ is the $l=2$ component of the non-relativistic deuteron
  wave function generated with the Paris potential \cite{
    PhysRevC.21.861, Lacombe1981139} (not to scale).}
\label{fig:corfunc}
\end{figure*}

The $g_{c}(r_{12})$ quantifies how strongly two point-like nucleons
treated as quasi-particles, are spatially correlated when they are a
distance $r_{12}$ apart. The $g_{c}(r_{12})$ gives rise to local
density fluctuations about the mean-field predictions from the
reference state $ \mid \Psi ^{MF} _A \ \rangle $.  The GD
$g_c(r_{12})$ (computed for nuclear matter) from
Fig.~\ref{fig:corfunc} is not very different from the one for
monoatomic molecules in a liquid. Indeed, for $r_{12} \rightarrow 0 $
one has that the GD $g_{c} \left( r_{12} \right) \rightarrow 1 $ which
reflects the fact that nucleons have a finite size (or, in other words
they are subject to a nucleon-nucleon interaction with a hard core).
For
values of $r_{12}$ which are larger than a few times the diameter of a
nucleon, the $g_{c} \left(r_{12} \right) \rightarrow 0 $. From this we
conclude that the fluctuations from the MF densities are confined to
short internucleon distances.  Therefore, the 2N SRC are a highly
local property and are insensitive to the properties of the other
surrounding nucleons. This is the fundamental reason why SRC can be
considered as ``universal'' \cite{Feldmeier:2011qy}. Whereas a large
model dependence for the $g_{c}$ is observed, the $f_{t \tau}$ seems
to be much better constrained. We have added the squared $D$-wave
component of the deuteron wave function $\Psi_{D} \left( k_{12}
\right)$ in Fig.~\ref{fig:corfunc}. Obviously, the momentum dependence
of $\mid f_{t \tau} \left( k_{12} \right) \mid ^{2}$ and the deuteron
momentum distribution $n_{D} \equiv \mid \Psi_{D} \left( k_{12}
\right) \mid ^{2}$ are highly similar.

The effect of the correlation functions on the momentum distributions
can be roughly estimated from their squared Fourier transforms. The
effect of the tensor correlation function is largest for moderate
relative momenta $ \left( 100 \lesssim k_{12} \lesssim 500
\right)$~MeV. For very large $k_{12}$, the $g_{c}$ is the dominant
contribution. The harder the $g_c(r_{12})$ the stronger the effect of
correlations. We stress that in the plane-wave impulse approximation,
the SRC contribution to the $(e,e'pp)$ cross section is proportional
to $\left|g_{c} \left( k_{12}\right)\right|^{2}$
\cite{Ryckebusch:1996wc}.

\begin{figure}
\includegraphics[viewport=0 0 346 246, clip, width=\columnwidth]{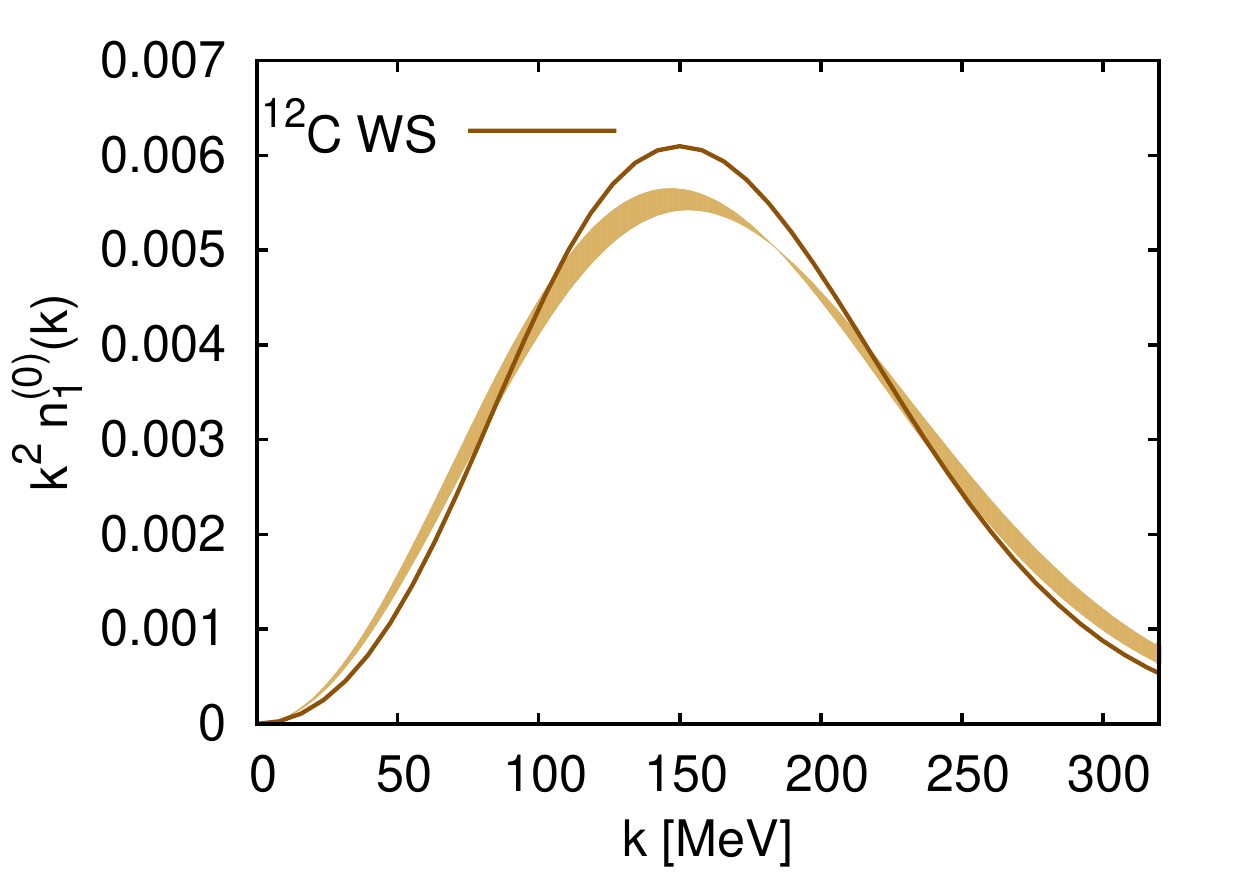}
\includegraphics[viewport=0 0 346 246, clip, width=\columnwidth]{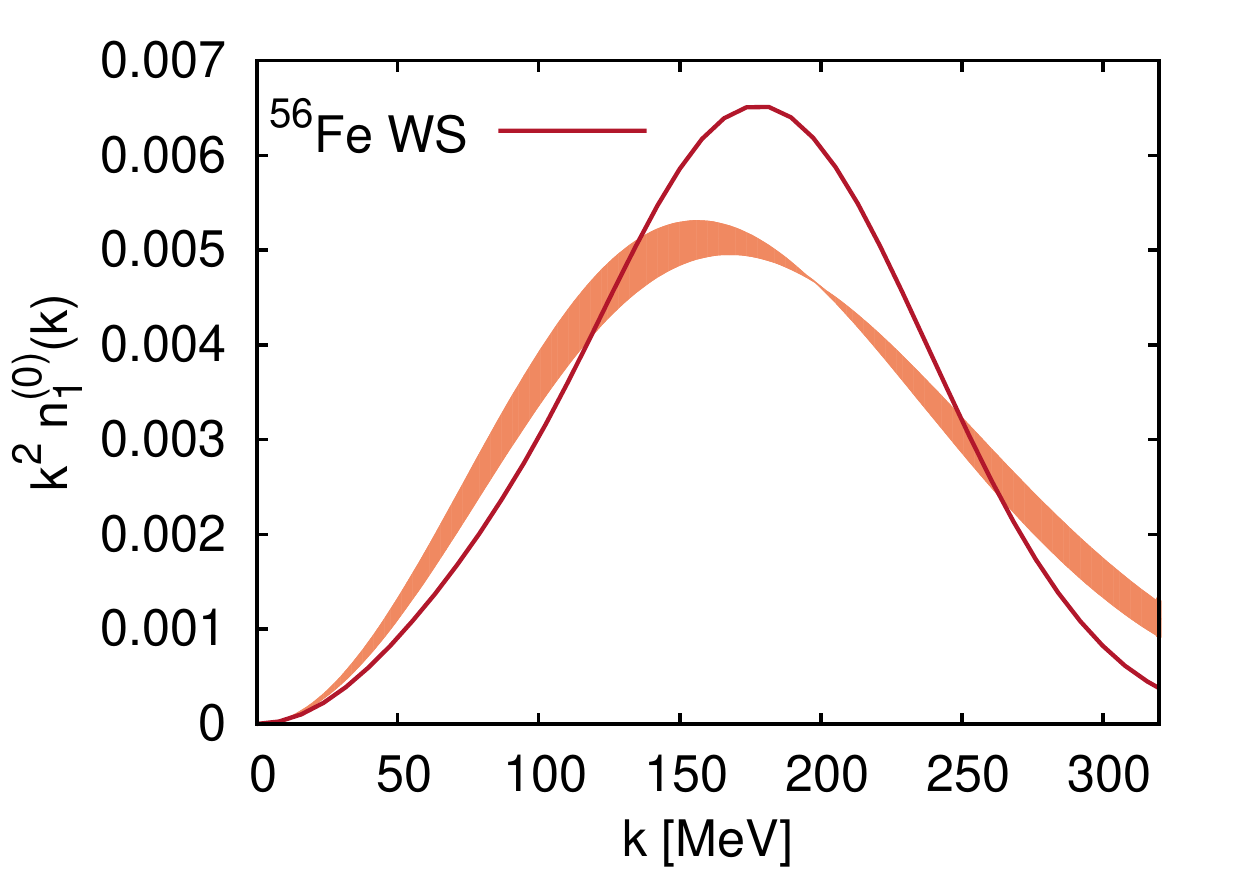}
\includegraphics[viewport=0 0 346 246, clip, width=\columnwidth]{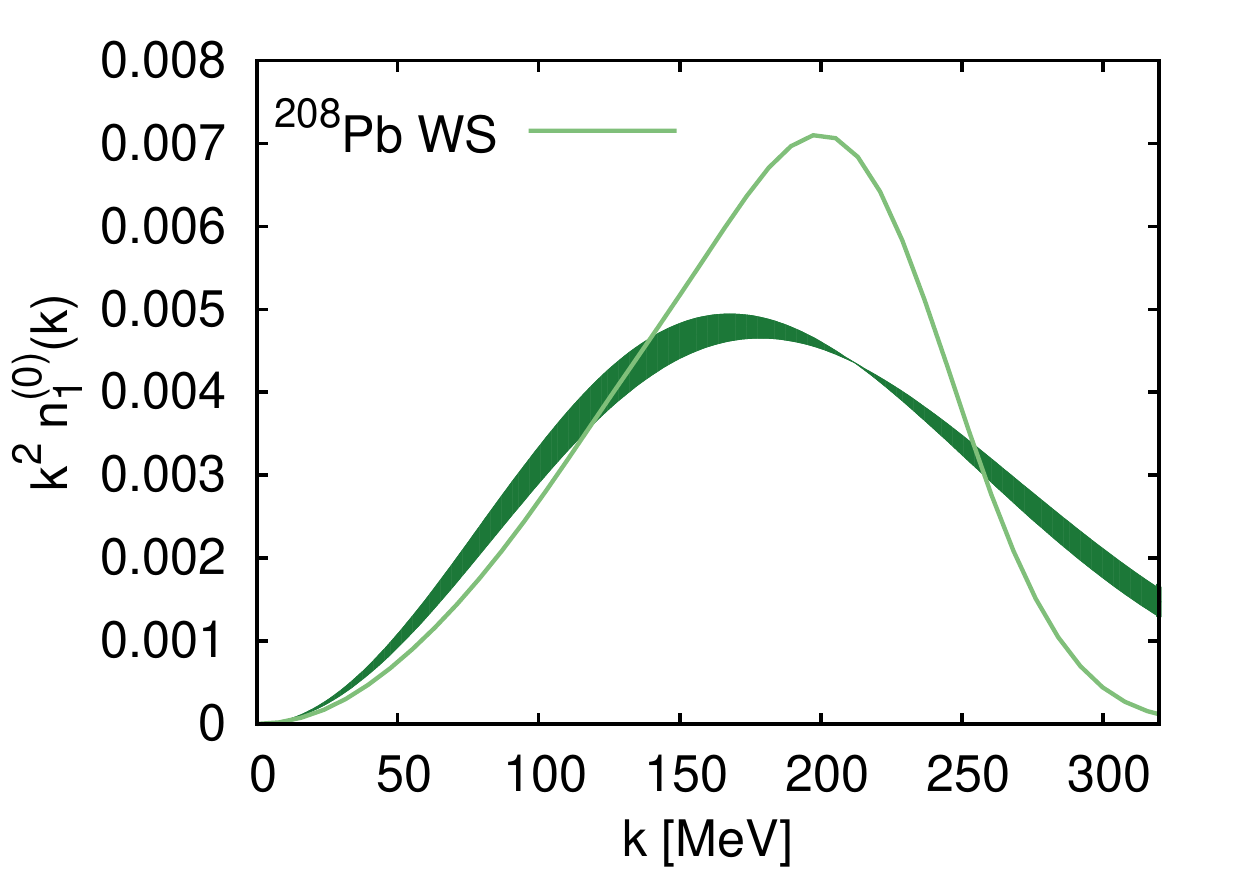}
\caption{(color online). The computed $k ^{2}n^{(0)}_1(k)$ versus $k$
  for the nuclei $^{12}$C, $^{56}$Fe and $^{208}$Pb and a Boltzmann
  fit including the error bars. The extracted values of $k T$ are
  12.0$\pm 0.5$~MeV (C), 14$\pm 1$~MeV (Fe), and 16$\pm 1$~MeV (Pb).
  The calculations are performed with WS single-particle states. The
  adopted normalization convention is $\int d k \; k ^{2} \;
  n^{(0)}_1(k) = 1$.}
\label{fig:n1kws}
\end{figure}

After introducing the wave functions of Eq.~(\ref{eq:realwf}), the
one-body and two-body momentum distributions of
Eqs.~(\ref{eq:momonebody}) and (\ref{eq:mom1}) can be written as
\begin{eqnarray}
P_{1}  \left( \vec{k} \right) 
& = &
P^{(0)}_1  \left( \vec{k} \right)  
+  P^{(1)}_1  \left( \vec{k} \right)   \; ,
\\
P_2 \left( \vec{k}_{12}, \vec{P}_{12} \right) & = &
P^{(0)}_2 \left( \vec{k}_{12}, \vec{P}_{12} \right) +
P^{(1)}_2 \left( \vec{k}_{12}, \vec{P}_{12} \right) \; .
\nonumber \\
& & 
\end{eqnarray}
The $ P^{(0)}_{1}  $ and $P^{(0)}_{2}$ are the mean-field parts and
are fully determined by the Slater determinant $ \mid \Psi _{A} ^{MF}
\ \rangle$. After inserting the expressions (\ref{eq:momonebody}) and
(\ref{eq:mom1b}) into the Eq.~(\ref{eq:n1k}) one obtains
\begin{eqnarray}
  n^{(0)}_1 \left( {k} \right) & = & \int d \Omega _{k} P^{(0)}_1 \left( \vec{k} \right) 
 = 
 \frac {2} { \pi } \sum _{n_h l_h j_h } \left( 2 j_ h + 1 \right) S_{n_h l_h j_h} 
\nonumber \\
& & \times \left( \int d r r^2 j_{l_{h}}(k r) \psi _{n_h l_h j_h} (r)   
\right) ^{2} \; ,
\label{eq:n1inMF}
\end{eqnarray}
where $j_{l} (r)$ is the spherical Bessel function of the first kind
and the sum extends over all occupied single-particle states. The $ 0
\le S_{n_h l_h j_h} \le 1$ is the occupation probability of the
corresponding single-particle state. The presence of short-range and
long-range correlations leads to occupation probabilities smaller than
one. With the adopted normalization convention of
Eq.~(\ref{eq:normap1k}) one typically obtains that
\begin{equation}
  \int d k k ^{2} n^{(0)}_1 \left( {k} \right) \approx 0.6 - 0.8 \; , 
\label{eq:normn0}
\end{equation} 
or, about $60-80 \%$ of the nucleons are mean-field like. We stress
that a considerable fraction of this depletion can be attributed to
long-range correlations, an effect which is not considered here.

The distribution $ k^{2} n^{(0)}_1 (k)$ as it can be computed from
Eq.~(\ref{eq:n1inMF}) is reminiscent for a phenomenon which is
confined to a certain scale, or, in other words, it is Gaussian
like. The typical scale is determined by the Fermi momentum $k_{F}
\approx 250~\textrm{MeV}$. This is illustrated in Fig.~\ref{fig:n1kws}
where we show the momentum dependence of the $ k ^{2} n^{(0)}_1 (k)$
for $^{12}$C, $^{56}$Fe and $^{208}$Pb as computed with Woods-Saxon
(WS) wave functions.  For the sake of curiosity we have fitted the
computed $ k ^{2} n^{(0)}_1 (k)$ with a Boltzmann distribution
\begin{equation}
\frac
{4 \pi} { \left( 2 \pi M_{N} k T \right) ^{3/2}}
 k ^{2} \exp - \frac { k ^{2} } { 2 M_{N}  k T} \; .
\label{eq:boltzmann}
\end{equation}
The results of the one-parameter fit are shown in
Fig.~\ref{fig:n1kws}. The fit is remarkably good for Carbon and gets
increasingly inaccurate with increasing mass number. From the fit of
the Boltzmann distribution we obtain $k T \approx 12$~MeV (C), $kT
\approx 14$~MeV (Fe), $kT \approx 16$~MeV (Pb). Accordingly, for the
IPM part of the momentum distribution, the typical energy exchange per
momentum degree-of-freedom $\frac {1} {2} k T $ is of the order of
6-8~MeV.

The correlated part $k ^2 n^{(1)}_1(k)$, on the other hand, is
reminiscent of the nucleus as a system of interdependent nucleons and
is obviously non-Gaussian.  In contrast to the mean-field part
$n^{(0)}_1$, the correlated part $n^{(1)}_1$ extends over ``all''
momentum scales.  Or, in other words the 2N, 3N, $\ldots$ correlations
generate a fat momentum tail to the $ n_{1} \left( k \right) $.  The
high momentum tails to $ n_{1} \left( k \right) $ have a very similar
form for all nuclei, including the deuteron, which alludes to some
universal character of SRC \cite{Feldmeier:2011qy}.

It has been theoretically predicted
\cite{frankfurt88,CiofidegliAtti:1995qe,janssen00} and experimentally confirmed in
semi-exclusive $A(e,e'p)$ measurements \cite{Iodice:2007mn} that the
major fraction of the $ n _{1}^{(1)} \left( k > k_{F} \right) $
strength is contained in very specific parts of the single-nucleon
removal energy-momentum phase space, namely those where the ejected
nucleon is part of a pair with high relative and small c.m. momentum.
This is the so-called ridge in the spectral function \cite{janssen00}
which reflects the fact that high-momentum nucleons in the one-body
momentum distribution are related to 2N dynamics with two nucleons
which are close and move back-to-back with approximately equal and
opposite momenta.

From recent calculations \cite{Alvioli:2011aa} of the two-body
momentum distributions in $^{3}$He and $^{4}$He the following
conclusions could be drawn. At high relative momenta and small
c.m. momenta, the c.m. and relative motion of the pair is decoupled,
an effect which is reminiscent of 2N SRC.  For the correlated pn pairs
the relative motion can be described by the high-momentum part of the
deuteron wave function. This suggests the following expression for the
correlated part of the pn two-body momentum distribution
\begin{eqnarray}
n^{(1)}_2 & & \left( 2 k_{F} \lesssim k_{12} , P_{12} \lesssim 150~\textrm{MeV} 
\right) 
\nonumber \\
& & \approx 
a_{pn} \left( A, Z \right) 
{n}_{D} \left( k_{12} \right) F ^{pn} \left( P_{12} \right)  \; ,
\label{eq:n2scaling}
\end{eqnarray}
where $ a_{pn} \left( A, Z \right) $ is a proportionality factor
related to the number of correlated proton-neutron pairs in the
nucleus $^{A}Z$ relative to the deuteron and $ {n}_{D} \left( k_{12}
\right) $ is the high-$k_{12}$ deuterium momentum distribution.
Further, the $ F ^{pn} \left( P_{12} \right)$ is the c.m. distribution
of the correlated pn pairs.  It corresponds with that part of
$P_{2}\left( P_{12} \right)$ of Eq.~(\ref{eq:comdis}) that stems from
pn pairs with a zero relative orbital angular momentum $l_{12}=0$ and
a total spin $S=1$.  The proposed scaling behavior
(\ref{eq:n2scaling}) can be attributed to the dominance of the tensor
correlations at medium relative momenta and the fact that $\mid f_{t
  \tau} \left( k_{12} > k_{F} \right) \mid ^{2} \sim \left| \Psi_{D}
  \left( k_{12} \right) \right| ^{2} $, two qualitative observations
which can made from Fig.~\ref{fig:corfunc}.

%
\subsection{Quantifying two-nucleon correlations}
\label{sec:II3}

We suggest that the significance of 2N correlations in a nucleus
$A(N,Z)$ is proportional to the number of relative $l_{12}=0$ states
\cite{Vanhalst:2011es}. There are experimental results supporting
this conjecture. First, in high-resolution $^{16}$O$(e,e'pp)^{14}$N
measurements performed at the electron accelerators in Amsterdam
\cite{ISI:000085387800006} and Mainz \cite{ISI:000222562800013}, the
quantum numbers of the target nucleus and the residual nucleus are
unambiguously determined. For the transitions to low-lying states in
the residual nucleus, the eightfold differential cross section for the
exclusive $(e,e'pp)$ reaction has been studied as a function of the
initial c.m. momentum $P_{12}$ of the proton-proton pair which is
involved in the reaction process. This has provided insight into the
quantum numbers of the pairs involved in the reaction process. We
denote by $\left| l_{12} \left( \vec{r}_{12} \right) , 
  \Lambda_{12} \left( \vec{R}_{12} \right) \right>$ the orbital wave
function corresponding with the relative and c.m. motion of a nucleon
pair.  For the ground-state (g.s.) to g.s. transition, for example,
\begin{equation}
 ^{16} \textrm{O} (0^+, g.s.) + e 
\longrightarrow ^{14} \textrm{C} (0^+, g.s.) + e ' + pp \; , 
\end{equation}
the active diproton resides in a state with quantum numbers $\left|
  l_{12} =0 , \Lambda_{12} =0 \right>$ at lower $P_{12}$ and $\left|
  l_{12} =1 , \Lambda_{12} =1 \right>$ at higher $P_{12}$. Two
independent calculations from the Pavia and Ghent groups have
demonstrated that the largest contributions from SRC to the eight-fold
cross section are confined to low $P_{12}$ values
\cite{ISI:000222562800013}. This provides direct evidence of pp
correlations being confined to $\left| l_{12} =0 , \Lambda_{12} =0
\right>$ pairs.  In that sense, the $^{16}$O$(e,e'pp)^{14}$N
measurements nicely confirmed the back-to-back picture of
SRC: diprotons are subject to SRC whenever
they happen to be close (or, in a relative $l_{12}=0$ state) and
moving back-back (or, in a state with $P_{12} \approx 0 $ which
corresponds with $\Lambda _{12}=0$).

High-resolution $(e,e'pn)$ measurements which have the potential to
access the pn correlations are very challenging
\cite{Middleton:2007rr}. Theoretical $(e,e'pn)$ calculations
\cite{janssen00,ryck00,Barbieri:2004xn} have predicted that the tensor
parts of the SRC are responsible for the fact that the correlated pn
strength is typically a factor of 10 bigger than the correlated pp
strength.  Calculations indicated that the tensor correlations are
strongest for pn pairs pairs with ``deuteron-like'' $\left| l_{12} =0,
  S=1\right>$ relative states \cite{ryck00,Barbieri:2004xn}.
Recently, the dominance of the pn correlations over pp and nn ones has
been experimentally confirmed \cite{Subedi:2008zz,
  PhysRevLett.105.222501}.

Accordingly, a reasonable estimate of the amount of correlated nucleon
pairs in $A(N,Z)$ is provided by the number of pairs in a 
$l_{12}=0$ state.  In order to determine that number for a given set
of single-particle states, one needs a coordinate transformation from
$(\vec{r}_1,\vec{r}_2)$ to $ \left(\vec{r}_{12}=\frac { \vec{r}_{1} -
    \vec{r}_2 } { \sqrt{2} }, \vec{R}_{12} = \frac {\vec{r}_{1} +
    \vec{r}_{2} } { \sqrt{2} } \right)$. For a harmonic oscillator
(HO) Hamiltonian this transformation can be done with the aid of
Moshinsky brackets \cite{BookHOMoshinsky}
\begin{eqnarray}
 & & \ket{n_1 l_1 \left( \vec{r}_1 \right) 
      n_2l_2 \left( \vec{r}_2 \right) ;LM_L}  =   
\sum_{n_{12}l_{12}N_{12}\Lambda_{12}} 
\nonumber \\  & & 
\braket{n_{12}l_{12}N_{12}\Lambda_{12};L}{n_1l_1n_2l_2;L}
\nonumber \\  & & \times
\ket{n_{12}l_{12} \left( \vec{r}_{12} \right)
  N _{12} \Lambda_{12} \left( \vec{R}_{12} \right) ;LM_L }\; . 
\label{eq_mosh}
\end{eqnarray}

We define the interchange operator for the spatial,
spin, and isospin coordinate as
\begin{equation}
\mathcal{P}_{12} = \mathcal{P}_{12} \left( \vec{r} _ 1 , \vec{r} _ 2 \right) 
        \mathcal{P}_{12} \left( \vec{\sigma} _ 1 , \vec{\sigma} _ 2 \right) 
        \mathcal{P}_{12} \left( \vec{\tau} _ 1 , \vec{\tau} _ 2 \right)  \; .
\label{eq:interchangeoperator}
\end{equation}
After introducing the spin and isospin degrees-of-freedom, in a
HO basis a normalized and antisymmetrized two-nucleon state reads
($\alpha _{i} \equiv (n_i l_i j_i t_i)$)
\begin{eqnarray}
& & 
\left| \alpha_{1}   \alpha _{2} ;  J M \right> _{na} 
  = \frac {1} { \sqrt { 2 \left( 1 + \delta _ {\alpha_{1} \alpha _{2}} \right)}}
 \nonumber \\ && \times 
\left( 1 - \mathcal{P}_{12} \right)
\left| \alpha_{1}   \left( \vec{r} _{1} \right) 
       \alpha _{2}  \left( \vec{r} _{2} \right)  
      ;  J M \right> 
\nonumber \\
& & =  
\sum _ {L M_L} \sum _{n_{12} l_{12}} \sum _{N_{12} \Lambda_{12}} 
\sum _{S M_S} \sum_{T M_T} 
\frac {1} { \sqrt { 2 \left( 1 + \delta _ {\alpha_{1} \alpha _{2}} \right)}}
\nonumber \\ & & 
\times \left[1-(-1)^{l_{12}+S+T}\right]
\nonumber \\ & & 
\times \braket{n_{12}l_{12}N_{12}\Lambda_{12};L}{n_1l_1n_2l_2;L}
\nonumber \\ & & 
\times 
\hat{j_1}\hat{j_2}\hat{L}\hat{S} 
 \begin{Bmatrix} l_1 & l_2 & L \\ \frac{1}{2} & \frac{1}{2} & S \\ j_1 & j_2 & J \end{Bmatrix} 
 \braket{L M_L S M_S}{J M} \; 
\nonumber \\ & & \times
\braket{\frac{1}{2} t_1 \frac{1}{2} t_2}{T M_T} 
\nonumber \\ & & 
\times \left|
\left[ {n_{12} l_{12} \left( \vec{r}_{12} \right)}, 
       {N_{12} \Lambda_{12} \left( \vec{R}_{12} \right) }
\right]LM_L, SM_S, TM_T \right> 
 \; ,
\nonumber \\ & & 
\label{eq:2bodynas}
\end{eqnarray}
where we have used the shorthand notation $\hat{j} \equiv \sqrt{2j+1}$.

\begin{figure}
 \includegraphics[width=0.9\columnwidth]{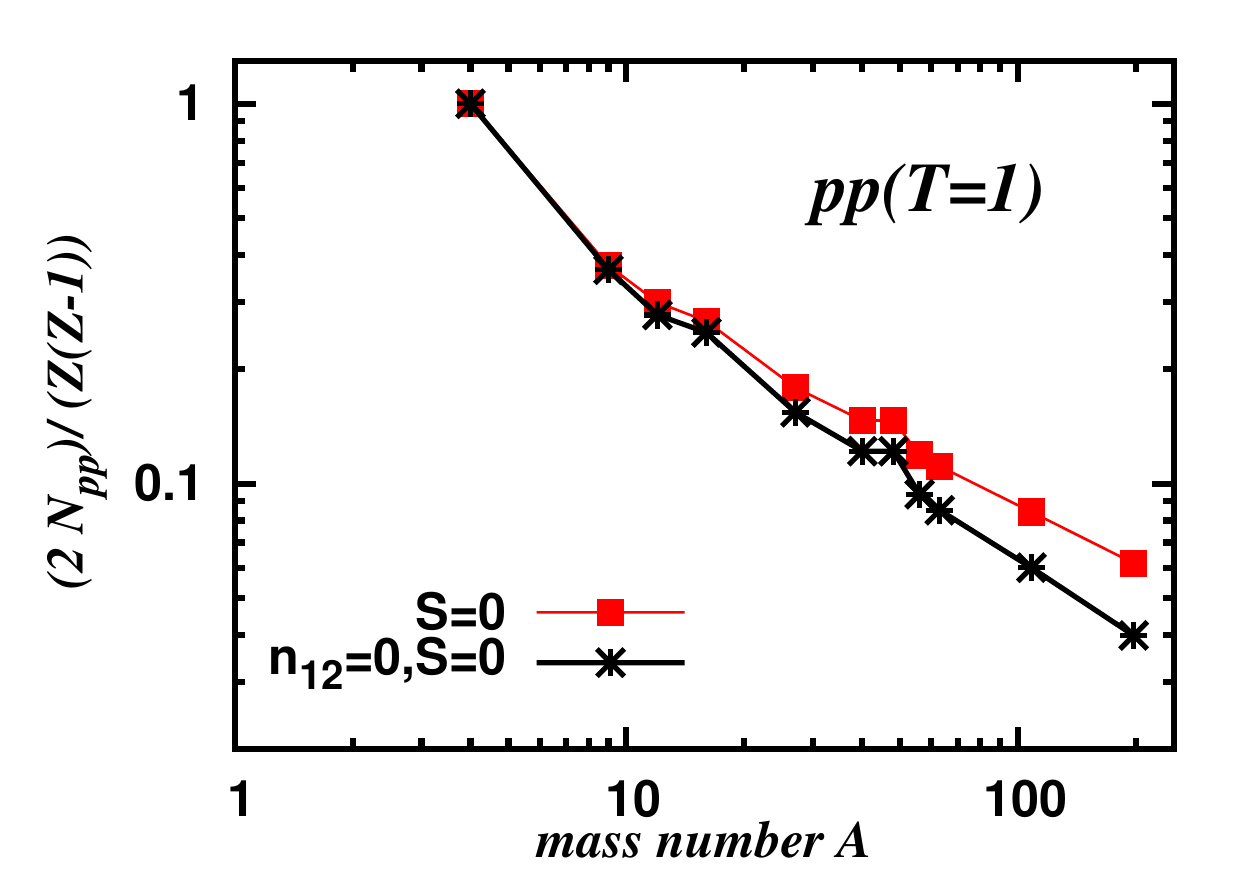}
 \includegraphics[width=0.9\columnwidth]{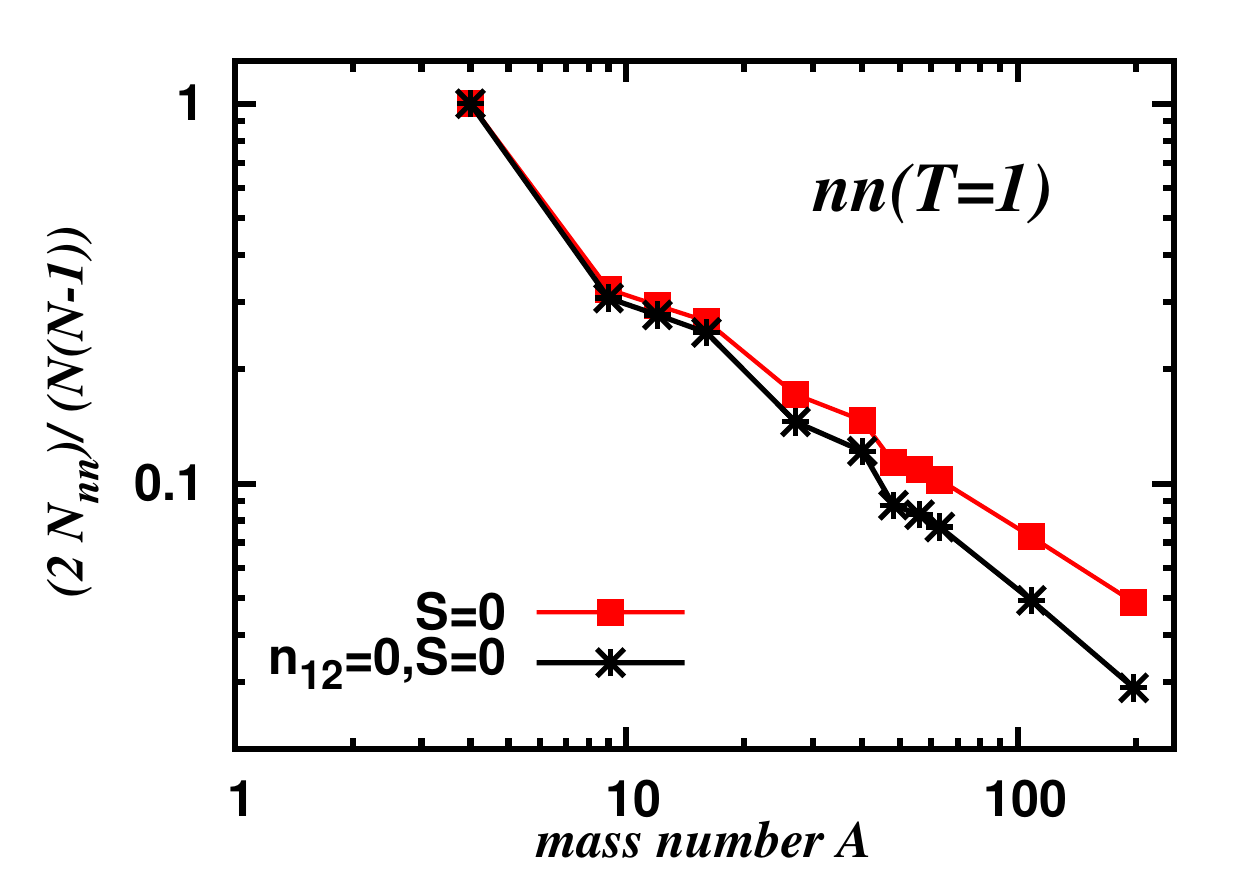}
  \includegraphics[width=0.9\columnwidth]{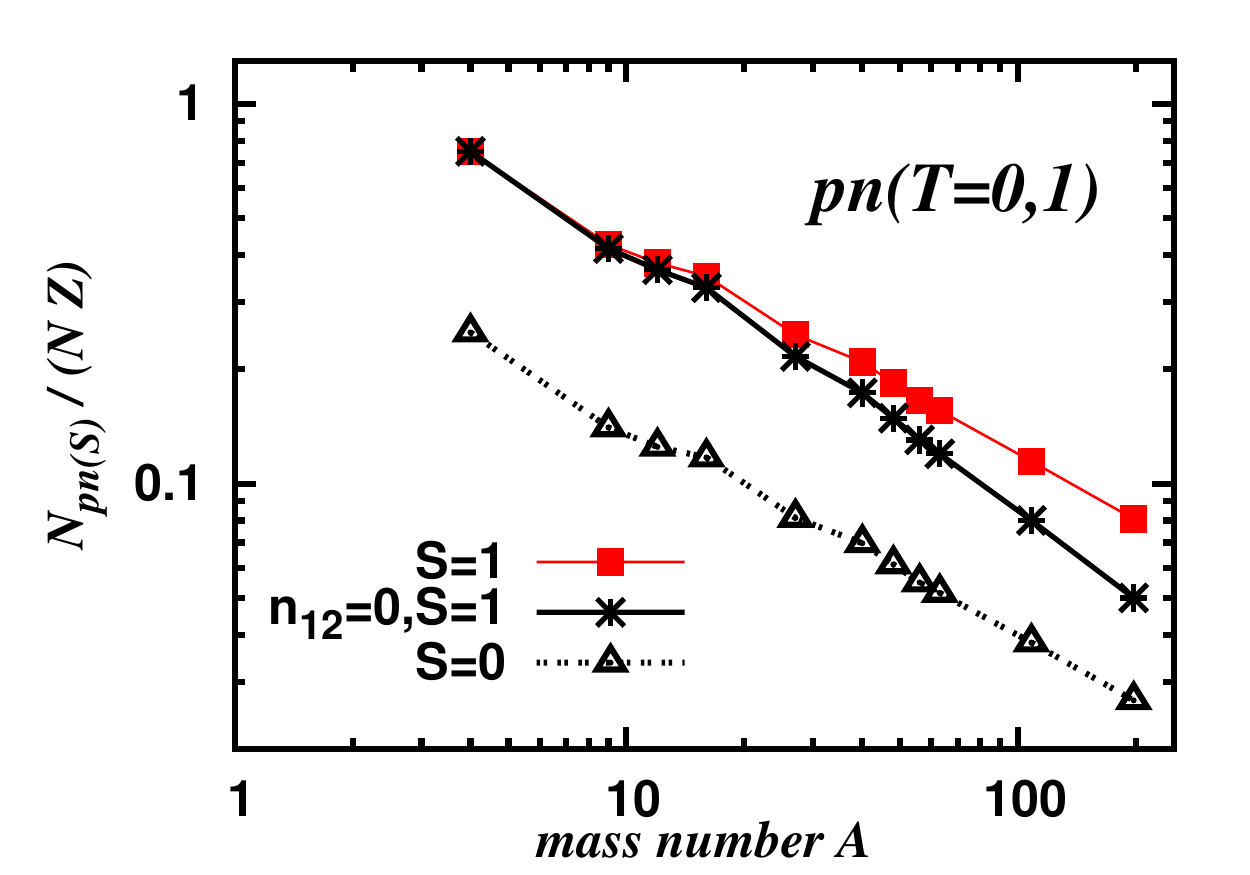}
  \caption{(color online). The computed values for $\frac {2} {Z(Z-1)}
    N_{pp} $, $\frac {2} {N(N-1)} N_{nn} $, and $\frac {1} {(NZ)}
    N_{pn(S)} $ which represent the predicted fraction of the pairs
    which are prone to SRC. The results are obtained for HO
    single-particle wave functions with $\hbar \omega (MeV) = 45. A ^{
      -\frac {1} {3}}- 25. A ^{ -\frac {2} {3}} $ and for the target
    nuclei $^{4}$He, $^9$Be, $ ^{12}$C, $ ^{16}$O, $ ^{27}$Al, $
    ^{40}$Ca, $ ^{48}$Ca, $ ^{56}$Fe, $^{63}$Cu, $ ^{108}$Ag, and $
    ^{197}$Au.}
\label{fig:spairnumbers}
\end{figure}

With the above conventions one has that the total amount of proton-neutron
pairs can be obtained from a sum over all pn pairs in the nuclear ground state 
\begin{equation}
\sum _{J M} \sum _{\alpha _{1}  \le \alpha _{F} ^{p}}
\sum _{\alpha _{2}  \le \alpha _F ^n} \;
  _{na} \left< \alpha_{1} \alpha _{2} ;  J M \right.
 \left| \alpha_{1} \alpha _{2} ;  J M \right> _{na} = N Z \; ,
\label{eq:sumrulepn}
\end{equation}
where $\alpha _F ^p$ and $\alpha _F^n$ denote the Fermi level for the
proton and neutron. Similar expressions hold for the number of
proton-proton and
neutron-neutron pairs 
\begin{eqnarray}
& & \frac {Z (Z -1)} {2}  =  \sum _{J M} \sum _{\alpha _{1}  \le \alpha _{F} ^{p}}
\sum _{\alpha _{2}  \le \alpha _F ^p}  \nonumber \\
& & \;
  _{na} \left< \alpha_{1} \alpha _{2} ;  J M \right.
 \left| \alpha_{1} \alpha _{2} ;  J M \right> _{na} \; ,  
\label{eq:sumrulepp}
\\
& & \frac {N (N -1)} {2}  =  \sum _{J M} \sum _{\alpha _{1}  \le \alpha _{F} ^{n}}
\sum _{\alpha _{2}  \le \alpha _F ^{n}} \nonumber \\
& &  \;
  _{na} \left< \alpha_{2} \alpha _{2} ;  J M \right.
 \left| \alpha_{1} \alpha _{2} ;  J M \right> _{na} \; .
\label{eq:sumrulenn}
\end{eqnarray}
Starting from the Eq.~(\ref{eq:2bodynas}) one can compute in a HO
single-particle basis how much a pair wave function with quantum numbers  
\begin{equation}
\left|\left[
n_{12}l_{12} \left( \vec{r}_{12} \right), N_{12}\Lambda_{12} \left( \vec{R}_{12}
    \right) \right] LM_L, SM_S, TM_T \right> 
\label{eq:pairwavefunction}
\end{equation}
contributes to the sum-rules of
Eqs.~(\ref{eq:sumrulepn},\ref{eq:sumrulepp},\ref{eq:sumrulenn}).  This
can also be done for any other basis $ \left| n l j m \right> $ of
non-relativistic single-particle states. In that case, the adopted procedure
involves an extra expansion of $ \left| n l j m \right> $ in a HO basis
\begin{eqnarray}
  & & \left| n l j m \right>  =  \sum _ { m_ {l} m _{s} } 
  \braket{l m_l \frac{1}{2} m_s}{j m} \psi _{n l j} (r) Y_{l m _{l} } \left( \Omega \right) 
\chi _{ \frac{1}{2} m_s} 
  \nonumber \\
& = & \sum _{n_{H}}  
\left( 
\int d r r ^{2} \phi_{n_{H} l} ^{*} (r)  \psi _{n l j} (r)
\right)
      \left| n_{H} l j m \right> \; , 
\label{eq:expansionHO}
\end{eqnarray}
where $ \phi_{n_{H} l} (r) $ are the radial HO wave functions. A
two-nucleon state can then be expressed in a HO basis for which the
Eq.~(\ref{eq:2bodynas}) can be used to determine the weight of the pair
wave functions of Eq.~(\ref{eq:pairwavefunction}).

The IPM pp pairs are mainly subject to the central SRC
which requires them to be close.  
This
implies that a reasonable estimate of the number of IPM pp pairs which
receive substantial corrections from the SRC is given by an expression
of the type
\begin{eqnarray}
& & N_{pp} (A,Z)  =  
\sum _{J M} \sum _{\alpha _{1}  \le \alpha _{F} ^{p}}
\sum _{\alpha _{2}  \le \alpha _F ^p}
\nonumber \\  & & 
 _{na} \left< \alpha_{1} \alpha _{2} ;  J M \right|
\mathcal{P}_{\vec{r}_{12}}^{l_{12}=0}
 \left| \alpha_{1} \alpha _{2} ;  J M \right> _{na} \; , 
\label{eq:project2NSRC}
\end{eqnarray}
where $ \mathcal{P}_{\vec{r}_{12}}^{l_{12}=0} $ is a projection
operator for two-nucleon relative states with $l_{12} =0 $.  A similar
expression to Eq.~(\ref{eq:project2NSRC}) holds for the nn pairs. For
the pn pairs it is important to discriminate between the triplet and
singlet spin states
\begin{eqnarray}
& & N_{pn(S)} (A,Z)  =  
\sum _{J M} \sum _{\alpha _{1}  \le \alpha _{F} ^{p}}
\sum _{\alpha _{2}  \le \alpha _F ^n}
\nonumber \\  & & 
 _{na} \left< \alpha_{1} \alpha _{2} ;  J M \right|
\mathcal{P}_{\vec{r}_{12}}^{l_{12}=0} 
\mathcal{P}_{\vec{\sigma}}^{S}
 \left| \alpha_{1} \alpha _{2} ;  J M \right> _{na} \; . 
\label{eq:projectnpSRC}
\end{eqnarray}


In Fig.~\ref{fig:spairnumbers} we display some computed results for
the $N_{pp}$, $N_{nn}$, and $N_{pn(S)}$ for 11 nuclei. The selection
of the nuclei is motivated by the availability of inclusive
electron-scattering data and covers the full mass range from Helium to
Gold. We have opted to display the results relative to the sum rule
values of the Eqs.~(\ref{eq:sumrulepn}) and (\ref{eq:sumrulepp}),
which allows one to interpret the results in terms of probabilities:
given an arbitrary pair wave function, what is the chance that it has
zero orbital relative momentum and a specific spin quantum number. In
a naive IPM picture for $^{4}$He, the pp pair is in a $ \left|
  l_{12}=0, S=0, T=1 \right>$ state. As this 2N configuration is prone
to central SRC effects, the corresponding probability is 1. The
physical interpretation is that for $^{4}$He ``all'' IPM pp-pair wave
function combinations receive corrections from SRC. For a medium-heavy
nucleus like $^{56}$Fe or $^{63}$Cu we find $\frac {N_{pp}} { \frac
  {Z(Z-1)} {2} } \approx 0.1$, which leads one to conclude that about
90\% of the IPM pp pair wave functions do not receive corrections from
central SRC.  For the heaviest nucleus considered here (Au) $ \frac {2
  N_{pp}} {Z(Z-1)} = 0.06 $, which means that only about 190 out of
the 3081 possible pp pair combinations are subject to SRC. 

Comparing the mass dependence of the pp and nn results of
Fig.~\ref{fig:spairnumbers} one observes similar trends. For the pn
results a softer decrease with increasing $A$ is predicted.  There are
about three times as many pn$(T=0)$ states than pn$(T=1)$ states with
$l_{12}=0$. This would be trivial in a system with only spin and
isospin degrees of freedom. In a system in which the kinetic energy
plays a role and in which there are spin-orbit couplings, we cannot
see any trivial reason why this should be the case. In this respect,
we wish to stress that for most nuclei discussed $N \ne Z$.  A
stronger criterion for selecting nucleon pairs at close proximity is
imposing $n_{12}=0$ in addition to $l_{12}=0$ and we have added also
those results to Fig.~\ref{fig:spairnumbers}. We find the results of
Fig.~\ref{fig:spairnumbers} robust in that the A dependence and
magnitudes are not very sensitive to the choices made with regard to
the single-particle wave functions. All the results of
Fig.~\ref{fig:spairnumbers} are displayed on a log-log plot and can be
reasonably fitted with a straight line, pointing towards a power-law
mass dependence $A^\alpha$ for the $N_{pp}$, $N_{pn}$ and
$N_{pn(S)}$.

%
\subsection{Quantifying three-nucleon correlations}
\label{sec:2D}

\begin{figure}
 \includegraphics[viewport=76 412 531 810, clip,
width=\columnwidth]{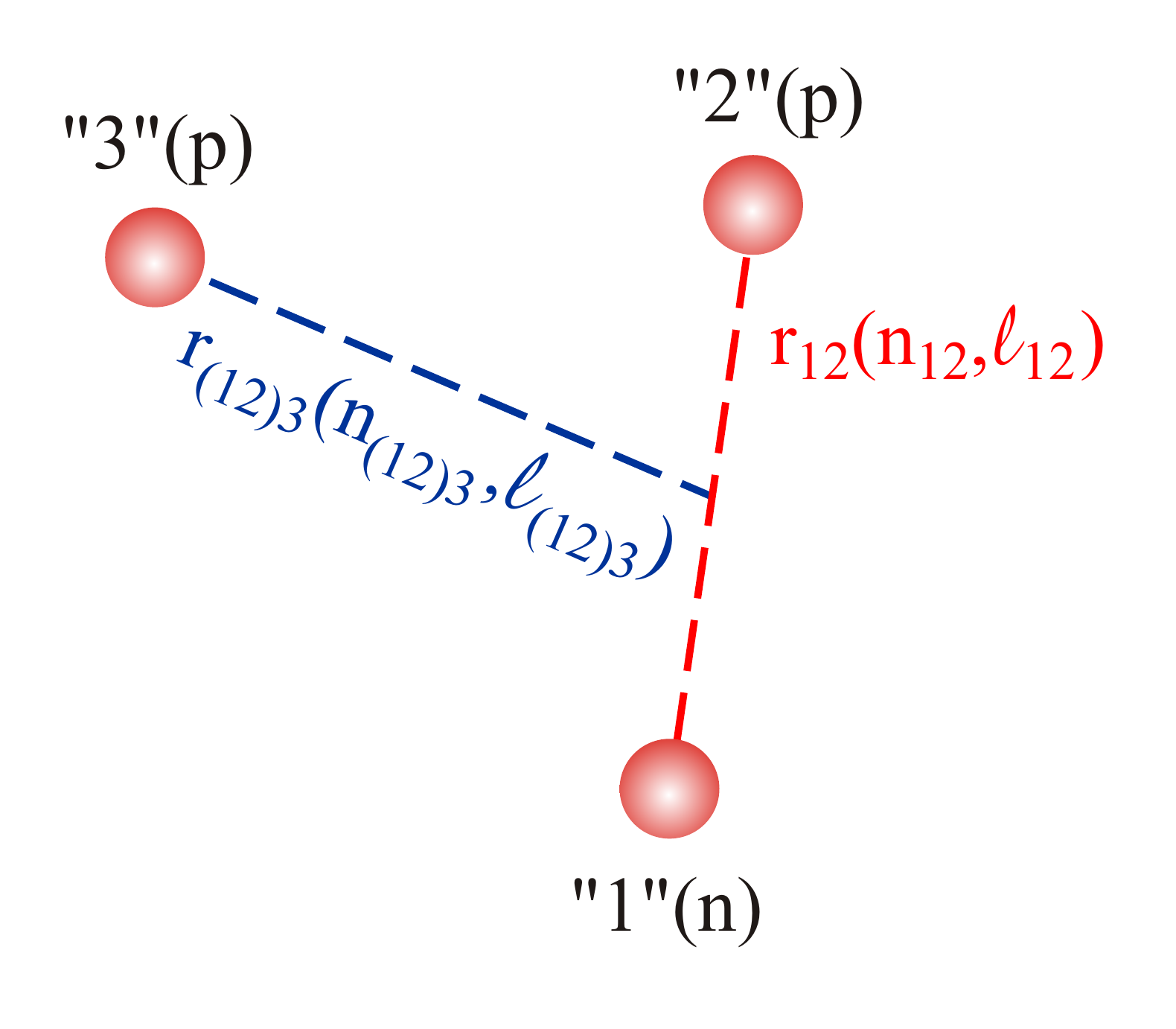}
\caption{(color online). Use of Jacobi coordinates in the ppn system.}
\label{fig:threebody}
\end{figure}

In order to quantify the magnitude of the 3N correlations
for an arbitrary $A(N,Z)$, we build on a recent paper
by Feldmeier et al. \cite{Feldmeier:2011qy}. There, it is pointed out
that 3N correlations can be induced without introducing genuine
three-body forces. In terms of the correlation operators of
Sect.~\ref{sec:IIB}, three-body correlations will naturally emerge in
cluster expansions by means of operators of the type
$g_c\left(r_{12}\right) g_c \left(r_{13}\right) $, $g_c\left(r_{12}\right) 
f_{t \tau} \left(r_{13}\right)$, $ \ldots \;$.

The strongest source of three-nucleon correlations is the tensor
correlation operator acting on the $(S=1,T=0)$ channel of the pn
states of Eq.~(\ref{eq:2bodynas}). We consider ppn configurations and
explain one possible mechanism to create a correlated state (see also
Fig.~\ref{fig:threebody}). In the uncorrelated wave function one has a
n(1)p(2) pair in a $\left| l_{12}=0 \; \; S_{12}=1 \; ; T_{12}=0
\right>$ state and a p(2)p(3) pair in a $\left| l_{23}=0 \; \;
  S_{23}=0 \; ; T_{23}=1 \right>$ state. Accordingly, both pairs are
in relative $l=0$ states. In Ref.~\cite{Feldmeier:2011qy} it is
explained that these two pairs can be brought into a correlated
three-nucleon status by flipping the spin of proton 2. In the
correlated part of the wave function one has an n(1)p(2) pair in a
$\left| l_{12}=2 \; \; S_{12}=1 \; ; T_{12}=0 \right> $ and an
p(2)p(3) pair in a $\left| l_{23}=1 \; \; S_{23}=1 \; ; T_{23}=1
\right> \; $ state. This configuration can be energetically favorable
through the presence of the strong tensor correlation in the pn
pair. Indeed, the energy gain through the tensor induced n(1)p(2)
correlation can compensate for the energy loss of breaking the pairing
in the p(2)p(3) pair. 

Given $A(N,Z)$ we propose to find all the antisymmetrized 3N states
with orbital quantum numbers 
\begin{equation}
(n_{12}=0 \; l_{12} =0 \; n_{(12)3}=0 \; l_{(12)3} =0) \; ,
\label{eq:Swavesfor3N}
\end{equation}
in the IPM wave function and identify them as the dominant contributors to 
3N SRCs.  This corresponds with seeking for those 3N
wave-function components where all three nucleons are ``close''. This
can be technically achieved by constructing antisymmetrized 3N states
starting from a MF Slater determinant, and performing a transformation
from the particle coordinates $\left(\vec{r}_1, \vec{r}_2, \vec{r}_3
\right)$ to the internal Jacobi coordinates $\left(\vec{r}_{12},
  \vec{r}_{(12)3} , \vec{R}_{123} \right)$
\begin{equation}
\vec{r}_{(12)3}  =  \frac {\vec{R}_{12} - \sqrt{2} \vec{r}_3 }{\sqrt{3}},
\hspace{0.1\columnwidth}
\vec{R}_{123}  =    \frac {\sqrt{2} \vec{R}_{12} + \vec{r}_3 }{\sqrt{3}} \; . 
\end{equation}
One readily finds for uncoupled three-nucleon states in a HO basis
\cite{BookHOMoshinsky}
\begin{eqnarray}
& & \left|n_1 l _1 m_{l_{1}} \left( \vec {r} _1 \right),
      n_2 l _2 m_{l_{2}} \left( \vec {r} _2 \right),
      n_3 l _3 m_{l_{3}}  \left( \vec {r} _3 \right) 
   \right>  =  \nonumber \\
&& \sum _ {L M_L} \sum _{n_{12} l_{12}} \sum _{N_{12} \Lambda_{12}} 
\sum _ {L_{1} M_{L_{1}}} \sum _{n_{(12)3} l_{(12)3}} 
\nonumber \\ && \times
\sum _{N_{123} \Lambda_{123}}
\sum _{m_{l_{12}} m_{\Lambda_{12}}} 
\sum _{m_{l_{(12)3}} m_{\Lambda_{123}}} \nonumber \\
& & \times   
\braket{l_{1} m_{l_{1}} l_{2} m_{l_{2}}}{L M_{L}}
\braket{l_{12} m_{l_{12}} \Lambda _{12} M_{\Lambda _{12}}}{L M_{L}}
\nonumber \\
& & \times \braket{ \Lambda _{12} M_{\Lambda _{12}} l_{c} m_{l_{c}}}{L_{1}
M_{L_{1}}}
\nonumber \\
& & \times
\braket{ l_{(12)3} m_{l_{(12)3}} \Lambda _{123} M_{\Lambda _{123}} }{L_{1}
M_{L_{1}}}
\nonumber \\
& & \times
\braket{n_{12}l_{12}N_{12}\Lambda_{12};L}{n_1l_1n_2l_2;L}
\nonumber \\
& & \times
\braket{n_{(12)3}l_{(12)3}N_{123}\Lambda_{123};L_1}{N_{12}\Lambda_{12}n_3l_3;L_1
} _{\beta}
\nonumber \\ 
& & \times
\left| n_{12} l _{12} m_{l _{12}} \left( \vec {r} _{12} \right) \right>
\left| n_{(12)3} l _{(12)3} m_{l _{(12)3}} \left( \vec {r} _{(12)3} \right)
\right>
\nonumber \\ 
& & \times
\left| N_{123} \Lambda _{123} M_{\Lambda _{123}} \left( \vec {R} _{123} \right)
\right>
\; ,
\label{eq:threebodytransform}
\end{eqnarray}
where we have adopted the notation $\braket{\ldots}{\ldots}_{\beta}$
for the Standard Transformation Brackets (STB) \cite{BookHOMoshinsky}.

\begin{figure}
\includegraphics[angle=-90, width=\columnwidth]{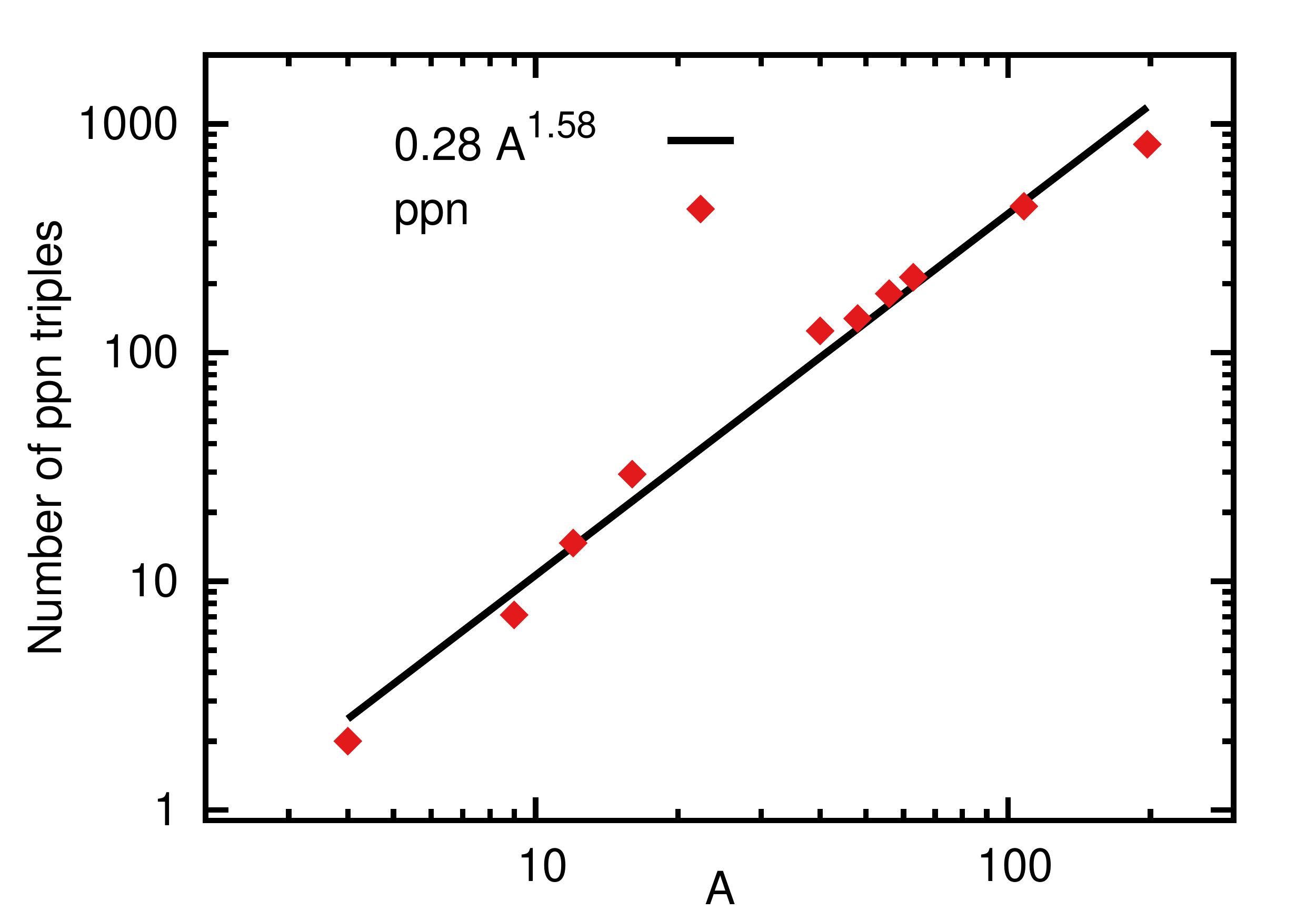}
\caption{(color online). The mass dependence of the amount of ppn triples with
  quantum numbers $\left| n_{12} =0 \; l_{12} =0, n_{(12)3} =0 \;
    l_{(12)3} =0 \right>$. The results can be reasonably fitted with a
  power law $0.28 A ^{1.58 \pm 0.20}$. 
  The results are obtained for HO single-particle wave functions with
  $\hbar \omega (MeV) = 45. A ^{ -\frac {1} {3}}- 25. A ^{ -\frac {2}
    {3}} $ and for the nuclei $^{4}$He, $ ^{9}$Be, $ ^{12}$C, $ ^{16}$O, $
  ^{40}$Ca, $ ^{48}$Ca, $ ^{56}$Fe, $ ^{63}$Cu, $ ^{108}$Ag, and $ ^{197}$Au.}
\label{fig:3NSRC}
\end{figure}

Antisymmetrized $(a)$ uncoupled three-nucleon states can be obtained from
the three-nucleon wave functions of Eq.~(\ref{eq:threebodytransform})
using the interchange operator of Eq.~(\ref{eq:interchangeoperator})
\begin{eqnarray}
& & \left| \alpha_{a}  m_{a} ,  \alpha _{b} m_{b} , \alpha _{c} m_{c} \right>
_{a} 
 =  
\nonumber \\
& & \left[ 1 - \mathcal{P}_{12} \right] 
\left| \alpha_{a}  m_a\left( \vec{r} _1 \right) ,
       \alpha_{b}  m_b \left( \vec{r} _2 \right) , 
       \alpha_{c} m_c \left( \vec{r} _3 \right)  \right> 
\nonumber \\
& + & \left[ 1 - \mathcal{P}_{12} \right] 
\left| \alpha_{b}  m_b\left( \vec{r} _1 \right) ,
       \alpha_{c}  m_c \left( \vec{r} _2 \right) , 
       \alpha_{a} m_a \left( \vec{r} _3 \right) \right> 
\nonumber \\
& + & 
\left[ 1 - \mathcal{P}_{12} \right] 
\left| \alpha_{c}  m_c \left( \vec{r} _1 \right) ,
       \alpha_{a}  m_a \left( \vec{r} _2 \right) ,
       \alpha_{b}  m_b \left( \vec{r} _3 \right) \right> \; .
\end{eqnarray}
The total number of ppn triples can now be expressed as 
\begin{eqnarray}
& & N \frac { Z (Z -1) } {2}  =   
\sum _{\alpha _{a}, \alpha _{b}  \le \alpha _F ^p}
 \sum _{\alpha _{c}  \le \alpha _F ^n}  \sum _{m_{a} m_{b} m_{c} }
\nonumber \\
 & & \; _{na} \left< \alpha_{a} m_{a} , \alpha _{b} m_{b} , \alpha _{c} m_{c}
\right.
 \left| \alpha_{a} m_{a} \alpha _{b} m_{b} \alpha _{c} m_{c} \right>
 _{na} 
   \; ,
\label{eq:pair44}
\end{eqnarray}
which allows for a stringent test of the analytical derivations and
their numerical implementation.  Along similar lines to those used to
derive the number of correlated 2N clusters in
Eq.~(\ref{eq:project2NSRC}), the number of ppn triples with the orbital
quantum numbers of Eq.~(\ref{eq:Swavesfor3N}) can be obtained from
\begin{eqnarray}
& & N_{ppn} (A,Z) = \sum _{\alpha _{a}, \alpha _{b}  \le \alpha _F ^p}
 \sum _{\alpha _{c}  \le \alpha _F ^n}  \sum _{m_{a} m_{b} m_{c} }
\nonumber \\
 & & \; _{na} \left< \alpha_{a} m_{a} , \alpha _{b} m_{b} , \alpha _{c} m_{c}
\right| 
\mathcal{P}_{\vec{r}_{12}}^{n_{12}=0,l_{12}=0}
\mathcal{P}_{\vec{r}_{(12)3}}^{n_{(12)3}=0,l_{(12)3}=0} \nonumber \\
& &  \left| \alpha_{a} m_{a} , \alpha _{b} m_{b} , \alpha _{c} m_{c} \right>
 _{na} \; . 
\label{eq:project3NSRC}
\end{eqnarray}
We associate the $N_{ppn} (A,Z) $ with the number of ppn SRC triples.
The $A$ dependence of $N_{ppn} (A,Z) $ is displayed in
Fig.~\ref{fig:3NSRC}.  There is striking linear correlation between
the logarithm of the mass number and the logarithm of the number ppn
triples which are close in the MF ground-state wave function.
%
\section{Results}
\label{sec:results}
In this section we discuss how our predictions for the number of
correlated 2N pairs and correlated 3N triples can be connected with
experimental results from inclusive electron scattering.

\subsection{Separation of the correlation and mean-field
  contributions}
\label{sec:resultsA}
We start with illustrating that the separation of the mean-field and
correlated contributions to the inclusive $A(e,e')$ cross sections is
feasible.  In order to achieve this, we use stylized features of the
$n_{1}(k)$ in Monte-Carlo (MC) simulations to illustrate that a
separation between the mean-field $n_{1} ^{(0)}(k)$ and the correlated
$n_{1} ^{(1)}(k)$ part can be made in the $A(e,e')$ signal.

\begin{figure}
 \centering
 \includegraphics[angle=-90,width=\columnwidth]{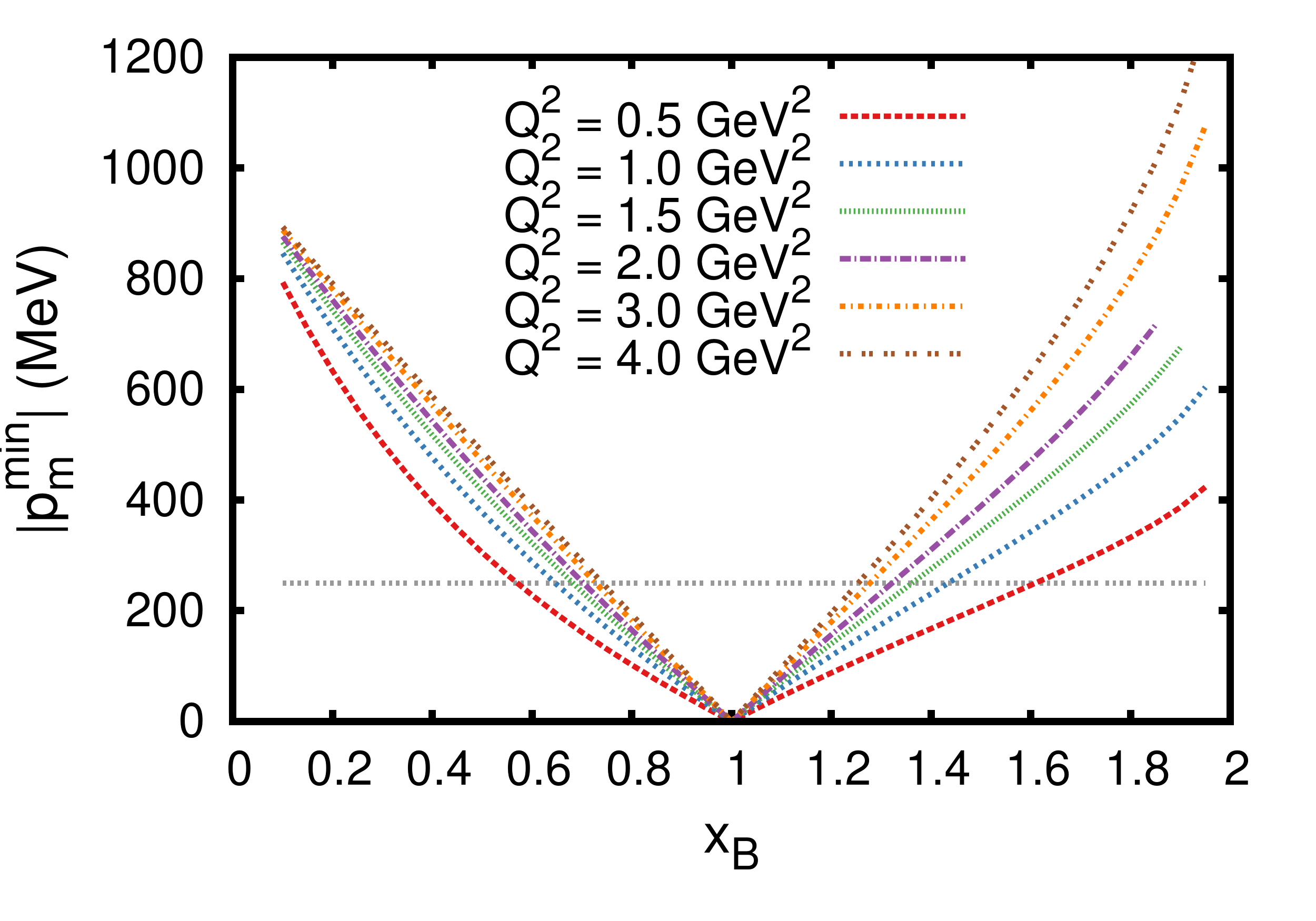}
 \caption{(Color online). Relation between the minimum of the missing
   momentum $|p_{m}^{min}|$ for the deuteron and $x_B$ at various values of the frour-momentum transfer $Q ^{2}$. }
 \label{fig:pm_min}
\end{figure}

We assume that quasi-elastic single-nucleon knockout $e + A \rightarrow
e' + (A-1) + N$ is the major source of $A(e,e')$ strength.  With
$q(\omega,\vec{q})$, $p_A(M_A, \vec{0})$,
$p_{A-1}(E_{A-1},\vec{p}_{A-1})$, $p_f(E_N, \vec{p}_f)$ we denote the
four-momenta of the virtual photon, of the target nucleus, of the
residual $A-1$ system, and of the ejected nucleon.  From
energy-momentum conservation
\begin{equation}
  q + p_A - p_{A-1} = p_f  , \label{eq:e-mcon1}
\end{equation}
 one can deduce for $A=2$ a relation between the minimum of the
 missing momentum $\vec{p}_m= \vec{p}_f-\vec{q}$ and the Bjorken
 scaling variable $x_B$ for fixed $Q^2$ \cite{Sargsian:2001ax}. The
 results are shown in Fig.~\ref{fig:pm_min}.  Obviously, for $Q^2 \geq
 1.5$~GeV$^{2}$ and $x_B>1.5$ one mainly probes nucleons with a
 momentum well above the Fermi momentum for the deuteron.  For finite
 nuclei the situation is more involving as $A-1$ represents an
 additional degree of freedom which can carry a fraction of the
 transferred four-momentum. We have performed MC simulations for a
 fixed energy of the impinging electron beam $\epsilon_i$ and a fixed
 electron scattering angle $\theta_e$.  The $p_{m}$ for a mean-field
 nucleon is drawn from the MF part $n_1^{(0)}(k)$ of $n_1(k)$.  For a
 correlated nucleon the $p_{m}$ is drawn from $n_1^{(1)}(k)$.
 Parameterizations for $n_1^{(0)}(k)$ and $n_1^{(1)}(k)$ are obtained
 from \cite{CiofidegliAtti:1995qe}
\begin{eqnarray}
  n_1^{(0)}(k) & = & A^{(0)} e^{-B^{(0)}k^2} [ 1 + \mathcal{O}(k^2) ],
\label{eq:n10} \\
  n^{(1)}_1(k) & = & A^{(1)} e^{-B^{(1)}k^2} + C^{(1)} e^{-D^{(1)} k^2},
\label{eq:n11}
\end{eqnarray}
where $A^{(0)}$, $B^{(0)}$, $A^{(1)}$, $B^{(1)}$, $C^{(1)}$ and
$D^{(1)}$ depend on $A$.

\begin{figure}
 \centering
 \includegraphics[width=\columnwidth]{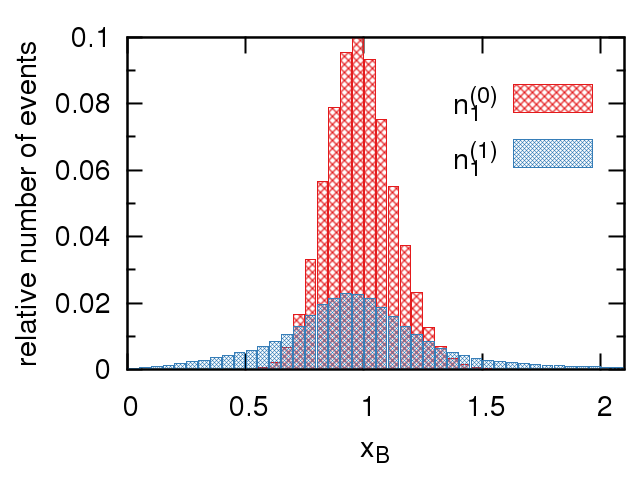}
 \caption{(Color online). Simulation of one-nucleon knockout in
   $^{12}$C with separated mean field and correlated momentum
   distribution.  The number of events is normalized as in
   Eq.~(\ref{eq:normn0}). The electron kinematics is determined by
   $\epsilon_i=\unit{5.766}{\giga\electronvolt}$ and
   $\theta_e=\unit{18}{\degree}$.}
 \label{fig:hist_n1}
\end{figure}

In Fig.~\ref{fig:hist_n1} we compare the $x_B$ distribution of
simulations for the mean-field and correlated part of one nucleon
knockout in $^{12}$C.  As stated in Eq.~(\ref{eq:normn0}), the number
of events is normalized as $\int \mathrm{d}k\; k^2 n_1^{(0)} (k) = 0.7$.
For $x_B> 1.5$, the events originate almost uniquely from
$n^{(1)}_1(k)$.  

%
\subsection{Two-body correlations}
\label{sec:resultsB}
Following the experimental observation \cite{PhysRevC.48.2451,
  PhysRevLett.96.082501,PhysRevLett.108.092502} that the ratio of the inclusive
electron scattering cross sections from a target nucleus $A$ and from
the deuteron $D$
\begin{equation}
\frac
{\sigma ^{A} \left(x_{B} , Q ^{2} \right) }
{\sigma ^{D} \left(x_{B} , Q ^{2} \right) } \; ,
\label{eq:defofratiodata}
\end{equation}
scales for $1.5 \lesssim x_{B} \lesssim 2$ and moderate $Q^{2}$, 
it has been suggested \cite{PhysRevC.48.2451} to parameterize the
$\sigma ^{A} $ in the following form
\begin{equation}
  \sigma ^{A} \left( 1.5 \lesssim x_{B} \lesssim 2, Q ^{2} \right) = \frac {A} {2} a_{2}
  \left( {A} / {D} \right) \sigma _{2} \left( A, x_{B} , Q ^2 \right)   \; , 
\label{eq:def_a2}
\end{equation}
where $ \sigma _{2} \left( A, x_{B} , Q ^2 \right) $ is the effective
cross section for scattering from a correlated 2N cluster in nucleus
$A$. Assuming that $ \sigma _{2 } $ is some local function which does
not depend on the target nucleus $A$
\begin{equation}
\sigma _{2} \left( A, x_{B} , Q ^2 \right)  \approx
\sigma _{2}\left( A=2, x_{B} , Q ^2 \right) \approx
\sigma ^D\left( x_{B} , Q ^2 \right) , 
\label{eq:localfunction}
\end{equation}
one can rewrite Eq. (\ref{eq:def_a2}) as
\begin{equation}
a_2 \left(  {A}/ {D} \right) 
= \frac {2} {A} \frac 
{\sigma ^{A} \left( x_{B}, Q ^{2} \right)} 
{\sigma ^{D} \left( x_{B}, Q ^{2} \right)} \; \; \left(1.5 \lesssim x_{B} \lesssim 2 \right) \; .
\label{eq:a2}
\end{equation}
In this simplified reaction-model picture, which ignores amongst other
things the c.m. motion of pairs in finite nuclei, the quantity $\frac
{A} {2} a_{2} \left( {A} / {D} \right)$ can be connected with the
number of correlated pairs in the nucleus $A$.  Assuming that all pn
pairs contribute one would expect that for the relative amount of
correlated two-nucleon clusters $a_{2} \left( {A} / {D} \right) \sim
A$.  Based on the observed dominance of correlated pn pairs over pp
and nn pairs \cite{Subedi:2008zz}, and the universality of the
deuteron-like high momentum tail of the correlated two-body momentum
distribution (\ref{eq:n2scaling}), we suggest that the correlated pn
pairs contributing to the $a_2(A/D)$, are predominantly $(T=0, S=1)$
pairs and that $ a_2 (A/D)$ is proportional to the quantity $N_{pn(S=1)}
(A,Z)$ defined in Eq.~(\ref{eq:projectnpSRC}).

In Ref.~\cite{Benhar:1995ph} the ratio of Eq.~
(\ref{eq:defofratiodata}) has been calculated with spectral functions
obtained from state-of-the-art nuclear matter calculations in the
local density approximation for the correlated part and $A(e,e'p)$
scattering data for the mean-field part
\cite{Benhar:1994hw,Sick:1994vj}.  The calculations suggested large
FSI effects, whilst the plane-wave calculations did not exhibit the
scaling present in the data at $1.5 \lesssim x_B$.  In
Refs.~\cite{Arrington:2011xs,Frankfurt:2008zv} it is argued that a
complete treatment of FSI in this kinematics needs to include
inelastic channels in the rescattering and that this cancels part of
the elastic FSI contribution included in Ref.~\cite{Benhar:1995ph}.
The results in Ref.~\cite{Benhar:1995ph} seem to refute the validity
of Eq.~(\ref{eq:def_a2}), which hinges on the assumption that the FSI
effects on correlated pairs in a nucleus are almost identical to those
in the deuteron in a high-momentum state. This requires that for $1.5
\lesssim x_B$ the FSI is primarily in the correlated pair and that the
remaining $A-2$ nucleons act as spectators.  Such a behavior is
suggested by the calculation of the quasi-elastic cross sections in
Ref.~\cite{CiofidegliAtti:1994ys} and by a space-time analysis of the
nuclear FSI at $x_B>1$ carried out in Ref.~\cite{Frankfurt:2008zv}
where it is stressed that the reinteraction distances are $\lesssim
1~$fm, supporting the idea that the first rescattering should be very
similar to FSI in the deuteron (see a recent discussion in
Ref.~\cite{Arrington:2011xs}). Therefore the assumption of
Eq.~(\ref{eq:def_a2}) seems a reasonable one for light nuclei where
the amount of rescatterings is of the order of 1.  For medium-heavy
and heavy nuclei, the average amount of rescatterings is larger than 1
and it has to be verified if the assumption still holds.  The
settlement and clarification of all the cited issues related to the
role of FSI in inclusive reactions requires further studies with a
full reaction model.

In a finite nucleus correlated pairs can have a non-zero c.m. momentum.
This c.m. motion is a correction factor when connecting the measured
$a_2(A/D)$ to the number of correlated pn pairs $N_{pn(S=1)}(A,Z)$.
We aim to provide an estimate for
this correction factor.  Therefore, we consider the two-nucleon
knockout reaction $e + A \rightarrow e' + (A-2) + N + N$ following the
break-up of a correlated 2N cluster.  For an inclusive cross section,
the tensor correlated pn$(S=1)$ pairs dominate the signal \cite{
  Piasetzky:2006ai, Subedi:2008zz, PhysRevLett.105.222501}.

As pointed out in Refs.~\cite{Frankfurt:2008zv, Ryckebusch:1996wc},
the cross section for the exclusive $(e,e'NN)$ reaction can be written
in a factorized form as
\begin{equation} 
\sigma^A(e, e' NN) = K F ^{NN}
(P_{12}) \sigma_{eNN} (k_{12} )\,, \label{eq:scalingfor2N}
\end{equation} 
where $P_{12} (k_{12} )$ is the c.m. (relative) momentum of the
correlated pair on which the photoabsorption takes place and $K$ is a
kinematic factor. The above expression is valid in the plane-wave and
spectator approximation for electron scattering on a pair with zero
relative orbital momentum.  The $\sigma_{eNN}$ stands for the
elementary cross section for electron scattering from a correlated 2N
pair with relative momentum $k_{12}$.  The $\sigma_{eNN}$ contains the
Fourier-transformed correlation functions $g_c(k_{12})$ and
$f_{t\tau}(k_{12})$.  An analytic expression for $\sigma_{epp}$ is
contained in Ref.~\cite{Ryckebusch:1996wc} and has been tested against
data in Ref.~\cite{Blomqvist:1998gq}.

As argued above, in order to link the exclusive cross section of
Eq.~(\ref{eq:scalingfor2N}) to the inclusive ones contained in the
Eq.~(\ref{eq:def_a2}) one assumes that $\sigma_{epn} \approx \sigma
_{eD} $ and one introduces a proportionality factor $N_{pn(S=1)}
(A,Z)$ which counts the number of correlated pn pairs in $A$. With the
scaling relation of Eq.~(\ref{eq:scalingfor2N}) for the $(e,e'pn)$
reaction, one can transform the ratio of Eq.~(\ref{eq:a2}) into a form
which accounts for the c.m. motion of the pair
 \begin{eqnarray}
   & & a_2(A/D) =  
   \frac{2}{A} \nonumber \\ 
& & \times \frac{ \int_{PS} \mathrm{d} \vec{k}_{12} \mathrm{d} \vec{P}_{12}
 N_{pn(S=1)}(A,Z)F ^{pn} (P_{12})  \sigma_{eD} \left( k_{12} \right)}
     { \int_{PS} d \vec{k}_{12} \sigma_{eD}( k_{12} ) }\;, 
\nonumber \\
   & \approx & \frac{2}{A} N_{pn(S=1)}(A,Z)  \int_{PS} d
 \vec{P}_{12} F ^{pn} \left( P_{12} \right)\;,
\label{eq:ourrelationfora2} 
\end{eqnarray}
where the integrations extend over those part of the c.m. momentum
phase (PS) included in the data. A basic assumption underlying the
above equation is that the factorization of Eq.~(\ref{eq:n2scaling})
approximately
holds.  The computed widths of the c.m. momentum distributions for the
correlated pn pairs contained in Table~1 indicate that the major
fraction of the pn pairs has $P_{12} \lesssim 150$~MeV which is within
the ranges for the validity of Eq.~(\ref{eq:n2scaling}).

\begin{figure}
 \centering
 \includegraphics[angle=-90,width=\columnwidth]{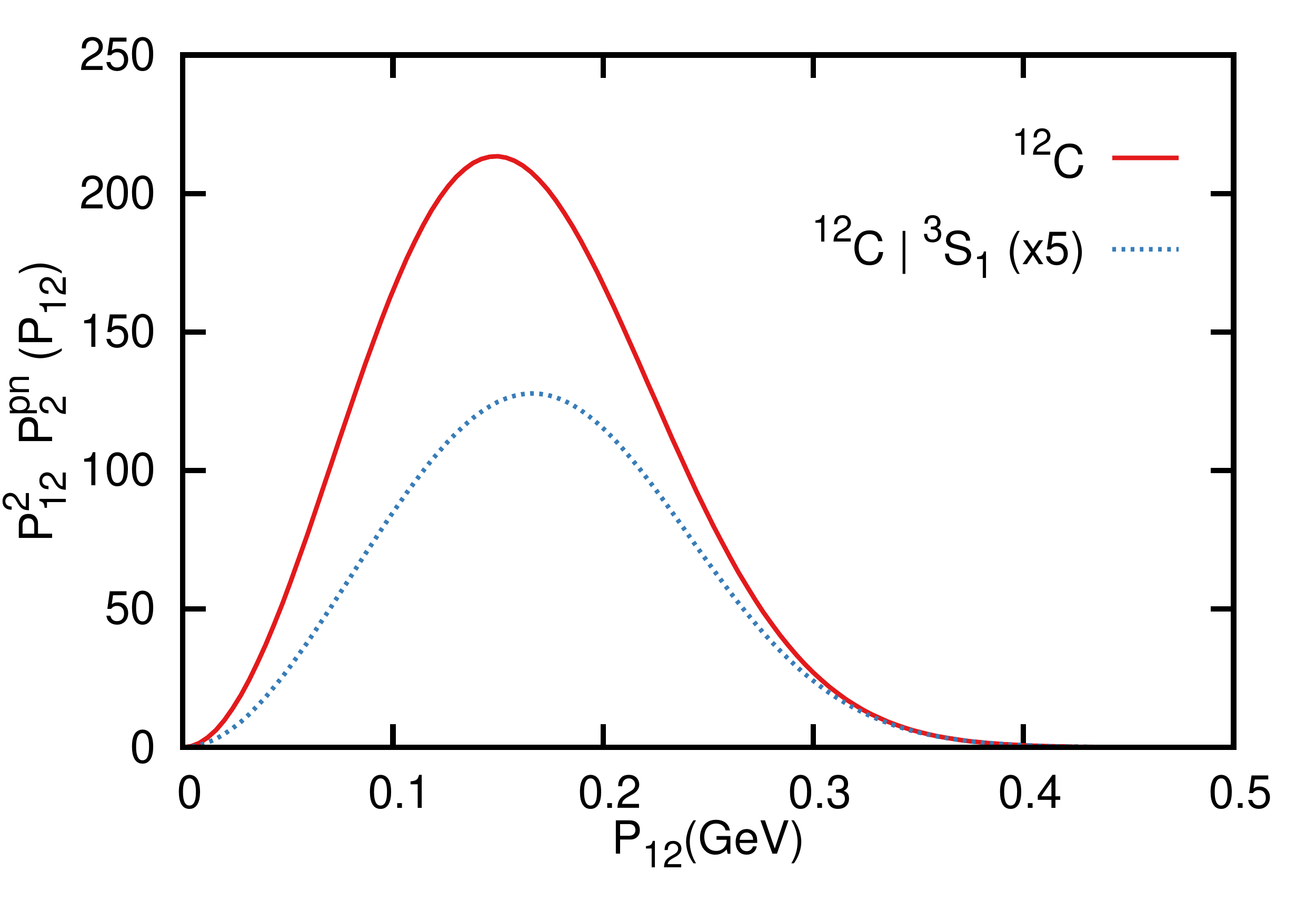}
 \caption{(Color online) The pn c.m. distribution $P_{12}^2 P_2^{pn}
   \left( P_{12} \right)$ and $P_{12}^2 P_2^{pn} \left( P_{12} | ^3S_1
   \right) = P_{12}^2 F^{pn} \left( P_{12} \right)$ for $^{12}$C.  The calculations
   are performed with HO single-particle states and adopt the
   normalization convention $\int d P_{12} \; P_{12}^2
   P_2^{pn} \left( P_{12} \right) = NZ $. }
 \label{fig:cm_distr}
\end{figure}

  In line with our assumption that the correlated
pairs are dominated by pn in a relative $^3S_1$ state,
$F ^{pn} \left( P_{12} \right)$ can be expressed as the
conditional two-body c.m. momentum distribution
\begin{equation}
  F ^{pn} \left( P_{12} \right) = P_2^{pn} \left( P_{12} | ^3S_1 \right) .
\end{equation}
Figure~\ref{fig:cm_distr} shows calculations for the $P_2^{pn} \left(
  P \right) $ and $P_2^{pn} \left( P | ^3S_1 \right)$ for
$^{12}$C. The c.m. distribution of correlated pn pairs ($F ^{pn}
(P_{12})$) can be well parameterized in terms of a Gaussian
distribution. The widths $\sigma_{c.m.}$ obtained from a Gaussian fit
to $P_2^{pn} \left( P_{12} | ^3S_1 \right)$ are given in
Table~\ref{tab:comcorrections}.

To estimate the c.m. correction factor we have performed
MC simulations of pn knockout with and without inclusion of
the c.m. motion.  This amounts to drawing the c.m. momentum from
$F ^{pn} (P_{12})= \delta( P_{12} )$ and $F ^{pn} (P_{12}) \sim e^{-
  \frac{P_{12}^2}{2\sigma_{c.m.}^2} }$, where $\sigma_{c.m.}$ is the
$A$ dependent width.  For $1.5 \leq x_B \leq 2$, the initial momentum
distribution of the correlated pair is given by correlated part of the
two-body momentum distribution $n_2^{(1)}( k_{12},P_{12})$.  The
Eq.~(\ref{eq:n2scaling}) states that the $n_2^{(1)} \left(k_{12},
  P_{12} \right)$ can be considered universal.
As illustrated in Fig.~\ref{fig:corfunc} one has $ {n}_D
\left( k_{12} \right) \sim \left| f_{t\tau}(k_{12}) \right|^{2}$.  As the relative
momentum distribution is
approximately proportional to the tensor correlation function, we draw
$k_{12}$ from the distribution $k_{12}^2 |f_{t\tau}(k_{12})| ^{2}$.
Energy conservation reads
\begin{equation}
  (q+ p_A- p_{A-2}- p_s)^2 = p_f^2= m_N^2,
  \label{eq:e-mcon2}
\end{equation}
where $q(\omega,\vec{q})$, $p_A(M_A,\vec{0})$ and
$p_{A-2}(E_{A-2},-(\vec{p}_s+\vec{p}_m))$ are the four-momenta of the
virtual photon, target nucleus and residual $A-2$ system,
respectively.  The virtual photon interacts with one of the nucleons,
resulting in a fast nucleon $p_f( E_f, \vec{p}_f)$ with $\vec{p}_f=
\vec{p}_m+ \vec{q}$ and a slow nucleon $p_s( E_s, \vec{p}_s )$.  With
the aid of Eq.~(\ref{eq:e-mcon2}), one can calculate the
$x_B$-distribution of the simulated events.  We apply the kinematics
of the Jefferson Lab (JLab) experiment E02-019 \cite{PhysRevLett.108.092502}: 
$\epsilon_i = \unit{5.766}{\giga\electronvolt}$ and $\theta_e=
\unit{18}{\degree}$.  The average $<Q^2>$ of the generated events
(including c.m. motion) in the $x_B$ region of interest is
2.7~GeV$^2$.  This value, which is $A$-independent, agrees with the
one quoted in Ref.~\cite{PhysRevLett.108.092502}.

\begin{figure*}
 \centering
 \includegraphics[width=\textwidth]{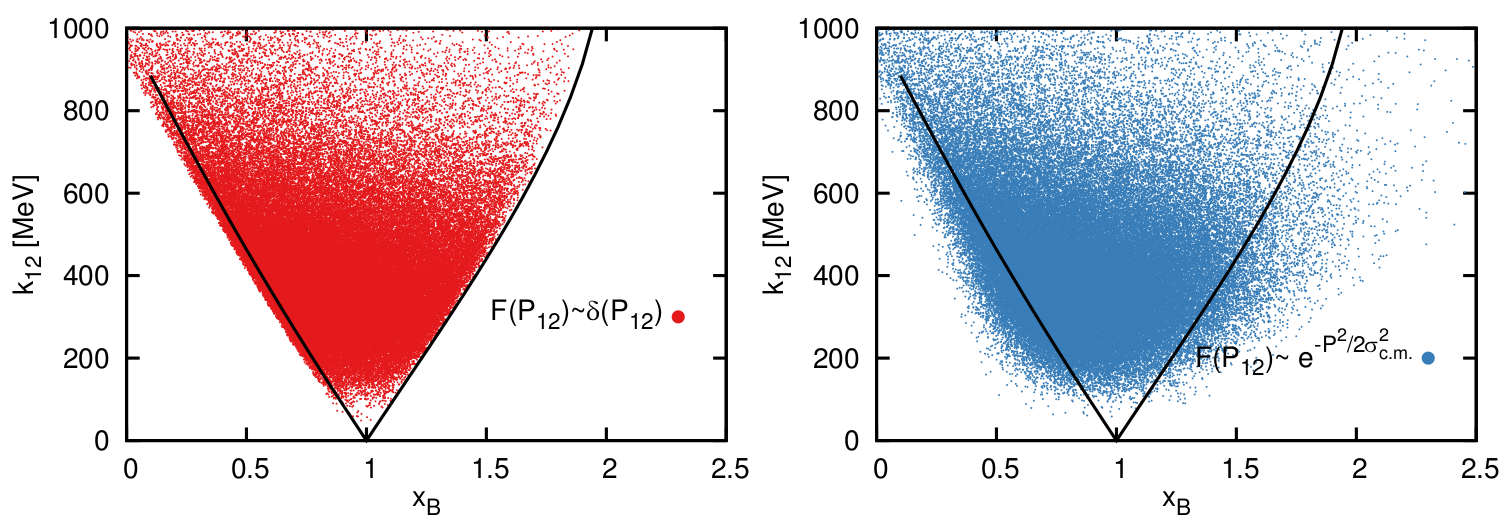}
 \caption{(Color online) The $k_{12}-x_B$ scatter plot of
   $^{12}$C$(e,e'p n)$ MC simulations with ($F(P_{12}) \sim e^{-
     \frac{P_{12}^2} { 2 \sigma_{c.m.}^2} }$ ) and without 
   ($F(P_{12}) \sim \delta(P_{12})$) inclusion of c.m. motion.  For the
   sake of comparison the solid line shows the minimum relative
   momentum $k_{12}^{min}$ for $Q^2=
   \unit{2.7}{\giga\electronvolt\squared}$ in the deuteron.}
 \label{fig:scatter}
\end{figure*}

\begin{figure}
 \centering
 \includegraphics[width=0.5\textwidth]{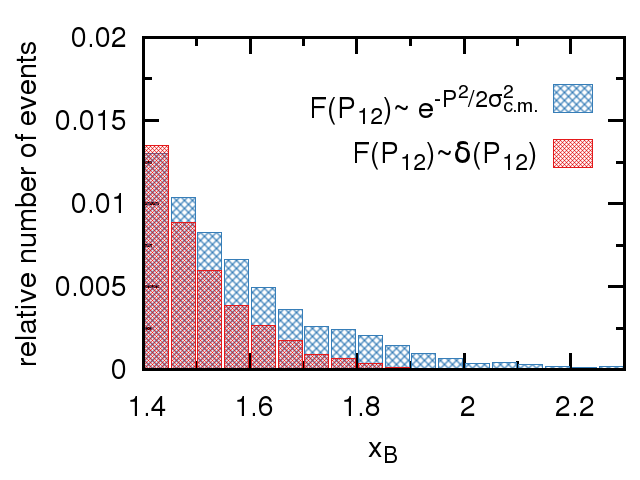}
 \caption{(Color online) Histogram of the $x_B$ distribution of
   $^{12}$C$(e,e'pn)$ MC simulations with ($F(P_{12}) \sim
   e^{- \frac{P_{12}^2} { 2 \sigma_{c.m.}^2} }$ ) and without ($F(P_{12})
   \sim \delta(P_{12})$) inclusion of c.m. motion. The kinematics is the one  
   of the JLab experiment E02-019 \cite{PhysRevLett.108.092502}: $\epsilon_i =
\unit{5.766}{\giga\electronvolt}$ and $\theta_e= \unit{18}{\degree}$.}
 \label{fig:hist_n2}
\end{figure}

The results of our simulations are summarized in
Figs.~\ref{fig:scatter} and \ref{fig:hist_n2}.  Fig.~\ref{fig:scatter}
shows the $x_B-k_{12}$ scatter plot of $10^6$ simulated events with
and without inclusion of c.m. motion for $^{12}$C.  In both situations
the mass difference between inital and final state causes a small shift 
to lower $x_B$ compared to the deuteron case.
Second, we observe considerable shifts in the distribution of
the events in the $(k_{12},x_{B})$ plane due to c.m. motion.  In
Fig.~\ref{fig:hist_n2}, one can observe how c.m. motion considerably
increases the number of events with $1.5 \leq x_B \leq 2$.  The impact
of the c.m. corrections increases with growing $x_B$.  Experimentally,
the $a_2(A/D)$ coefficient is determined by integrating data for 
$1.5 \leq x_B \leq 1.85$. We estimate the c.m. correction factor by the ratio
\begin{equation}
\frac {\textrm{\# simulated events with inclusion of c.m. motion}} 
      {\textrm{\# simulated events without inclusion of c.m. motion}} \; .
\label{eq:correctionfactor}
\end{equation}
in this $x_{B}$ region.  The resulting correction factor for several
nuclei is contained in Table~\ref{tab:comcorrections}. We performed
the simulations with the three different correlation functions $f_{t
  \tau}$ in Fig.~\ref{fig:corfunc}.  The dependence of the result on
the choice of correlation function is represented by the error of the
c.m. correction factor.

\begin{table}
\begin{tabular}{|l|c|c|c|}
\hline \hline
& A 		& $\sigma_{c.m.}$ & c.m. correction factor 	\\
\hline
& $^{12}$C   	& $115$ MeV	& $1.64 \pm 0.23$ 	\\ 
& $^{56}$Fe  	& $128$ MeV    	& $1.70 \pm 0.27$	\\ 
& $^{208}$Pb  	& $141$ MeV     & $1.71 \pm 0.29$	\\ 
\hline \hline 
\end{tabular}
\caption{The second column gives the width of the c.m. distribution of 
  correlated pn pairs. The third column provides the computed 
  c.m. correction factor.  
  The errors represent the dependence on the choice of correlation function.}
\label{tab:comcorrections}
\end{table}

\begin{figure}
 \includegraphics[width=\columnwidth]{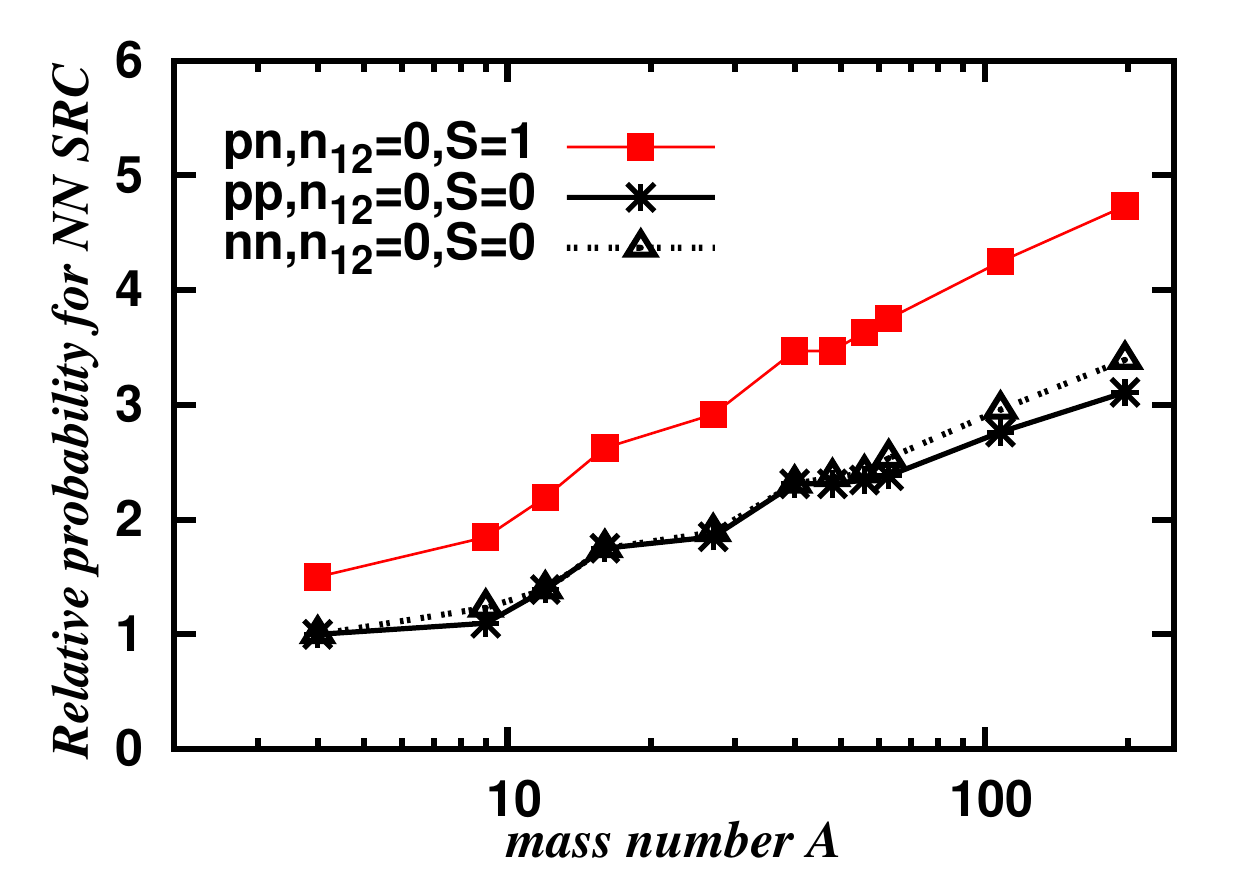}
 \caption{The mass dependence of the per nucleon probability for pn
   SRC relative to the deuterium. }
\label{fig:reltodeut}
\end{figure}

Fig.~\ref{fig:spairnumbers} quantifies the fraction of all possible pn
pairs which are prone to SRC relative to the total amount of possible
pn pair combinations. In our picture one has $N_{pn(S=1)}=1$ for
D. This means that we do interpret the $l_{12}=0$ component of the
deuteron wave function as the IPM part which receives large
corrections from tensor SRC. The per nucleon probability for a pn SRC
relative to the deuterium can be defined as
\begin{equation}
\frac {2} {N+Z} \frac {N_{pn(S=1)}(A,Z)}{N_{pn(S=1)}(A=2,Z=1)} = 
\frac {2} {A}  {N_{pn(S=1)}(A,Z)} 
\; .
\label{eq:pernucleonSRCrelativetod}
\end{equation}
Similar expressions hold for the per nucleon pp SRC and the per nucleon nn SRC
\begin{equation}
\frac {2} {Z}  {N_{pp(S=0)}(A,Z)} 
\quad \quad \frac {2} {N}  {N_{nn(S=0)}(A,Z)} \; . 
\end{equation}
The results of the per nucleon probabilities are collected in
Fig.~\ref{fig:reltodeut}.  Relative to $^{2}$H, the per nucleon
probability of pn SRC are 2.20, 3.63, 4.73 times larger for Carbon,
Iron, Gold. Along similar lines, relative to the ``free'' pp system the per
nucleon probability of pp SRC are 1.39, 2.34, 3.11 times larger for
Carbon, Iron, Gold.

\begin{figure}
 \centering
 \includegraphics[width=\columnwidth]{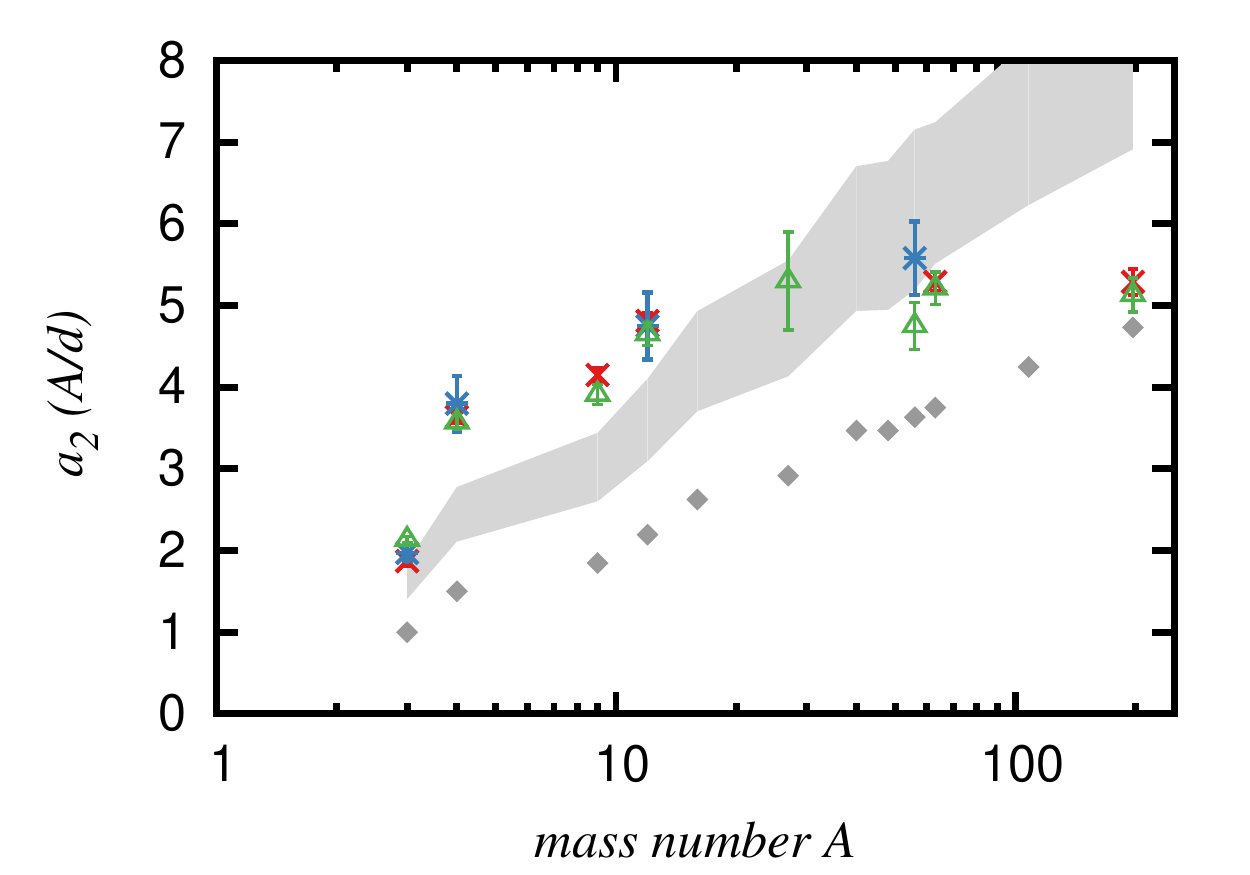}
 \caption{(Color online) The computed values for the $a_2 (A/D)$ for
   various nuclei. The data are from
   Refs.~\cite{PhysRevLett.96.082501,PhysRevLett.108.092502,Arrington:2012ax}.
   The shaded region is the prediction after correcting the computed
   values of $a_2(A/D)$ for the c.m. motion of the pair. The
   correction factor are determined by linear interpolation of the
   factors listed in Table \ref{tab:comcorrections}.  The width of
   the shaded area is determined by the error of the c.m. correction
   factors.}
 \label{fig:pairs4}
\end{figure}

In Fig.~\ref{fig:pairs4} we compare our predictions computed with the
aid of the Eq.~ (\ref{eq:ourrelationfora2}) with the extracted values
of $a_{2}(A/D)$. We have opted to correct the predicted $a_{2}$
coefficients and not the data for c.m. motion.  We stress that the
c.m. correction factor cannot be computed in a model-independent
fashion. For light nuclei our predictions tend to underestimate the
measured $a_{2}$. This could be attributed to the lack of long-range
clustering effects in the adopted wave functions. Indeed, it was
pointed out in Ref.~\cite{PhysRevC.83.035202} that the high-density
cluster components in the wave functions are an important source of
correlation effects beyond the mean-field approach.  For heavy nuclei
our predictions for the relative SRC probability per nucleon do not
saturate as much as the data seem to indicate.  
In Ref. \cite{Frankfurt:2008zv} the authors estimated the mass
dependence of $a_2$ by means of an expression of the type $a_2 \sim
\int d^3\vec{r} \rho^2_{\text{MF}}(\vec{r})$.  Using
Skyrme Hartree-Fock densities $\rho _{\text{MF}}(\vec{r})$ a power-law of
$A^{0.12}$ emerged. After normalizing to the measured value for
$a_2(^{12}$C$/D)$ the predicted power-low dependence agrees nicely
with the data.

We stress that final-state interactions (FSI) represent another source
of corrections which may induce an additional $A$-dependent correction
to the data.  FSI of the outgoing nucleons with the residual spectator
nucleons, could shift part of the signal strength out of the cuts
applied to the experimental phase space (or likewise move strength in)
and decrease (or increase) the measured cross section and the
corresponding $a_2$ coefficient.

\begin{figure}
\includegraphics[width=0.7\columnwidth,angle=-90]{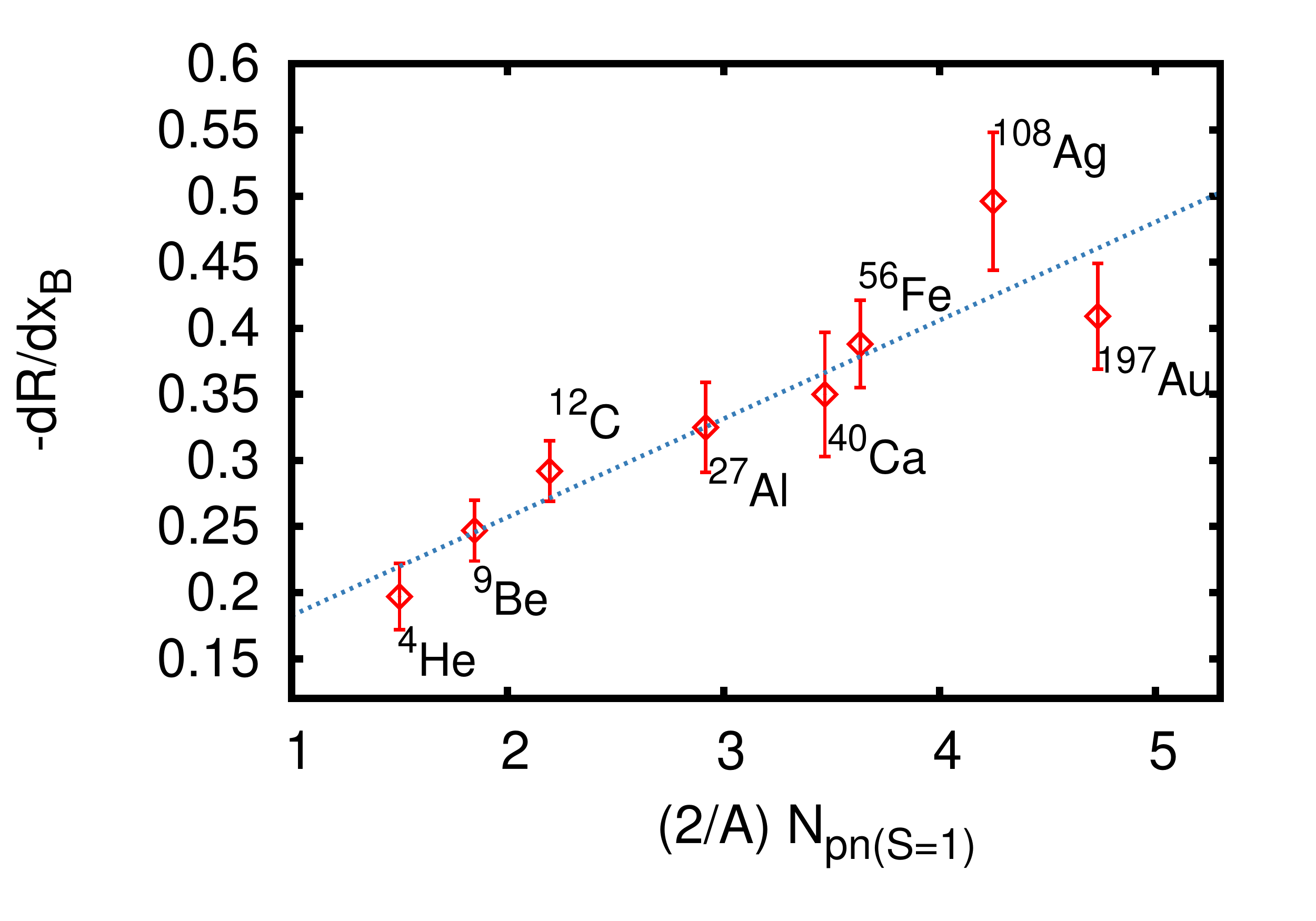}
\caption{The magnitude of the EMC effect versus the computed per
  nucleon number of correlated pn pairs. The data are from the
  analysis presented in
  Refs.~\cite{Seely:2009gt,PhysRevD.49.4348,Arrington:2012ax}. The
  fitted line obeys the equation $-\frac{d R}{d x_{B}} = (0.108 \pm
  0.028) + \frac {2} {A} N_{pn(S=1)} \cdot (0.074 \pm 0.010) $. }
\label{fig:EMCeffect}
\end{figure}

In Fig.~\ref{fig:EMCeffect} we display the magnitude of the EMC
effect, quantified by means of $- \frac {d R} {d x_{B}}$ versus our
predictions for the quantity $\frac{2} {A} N_{pn(S=1)}$ or, the "per
nucleon probability for pn SRC relative to the deuteron". We stress
that the numbers which one finds on the x-axis are the results of
parameter-free calculations. We consider the "per nucleon probability
for pn SRC relative to the deuteron" as a measure for the magnitude of
the proton-neutron SRC in a given nucleus. Obviously, there is a nice
linear relationship between the quantity which we propose as a per
nucleon measure for the magnitude of the SRC and the magnitude of the
EMC effect. 

%
\subsection{Three-body correlations}
\label{sec:resultsC}
The measurements of
Refs.~\cite{PhysRevLett.96.082501,PhysRevLett.108.092502} indicate that the
ratio of the inclusive cross sections
\begin{equation}
\frac
{\sigma ^{A} \left(x_{B} , Q ^{2} \right) }
{\sigma ^{^{3}\textrm{He}} \left(x_{B} , Q ^{2} \right) } \; ,
\label{eq:a3deriv1}
\end{equation}
approximately scales for $2.25 \lesssim x_{B} \lesssim 3.0$. Along
similar lines as those used in quantifying the 2N SRC in
Sec.~\ref{sec:resultsB}, it has been suggested
\cite{PhysRevLett.96.082501} to parameterize the inclusive $A(e,e')$
cross section in the following form
\begin{equation}
\sigma ^{A} \left( 2.25 \lesssim x_{B} \lesssim 3, Q ^{2} \right) = 
\frac {A} {3} a_{3} \left(  {A} / {^{3}\textrm{He}} \right)  {\sigma _{3} \left( x_{B} ,
Q ^2 \right) }  \; ,
\label{eq:a3deriv2}
\end{equation}
where $ {\sigma _{3} \left( x_{B} , Q ^2 \right) } $ is the cross
section for scattering from a correlated 3N cluster which is once
again assumed to be $A$ independent.  Inserting
Eq.~(\ref{eq:a3deriv2}) into Eq.~(\ref{eq:a3deriv1}), one obtains
\begin{equation}
a_3 \left(  {A}/ ^{3}\textrm{He} \right) 
= \frac {3} {A} \frac 
{\sigma ^{A} \left( x_{B}, Q ^{2} \right)} 
{\sigma ^{^{3}\textrm{He}} \left( x_{B}, Q ^{2} \right)} \; \; \left(2.25 \lesssim x_{B} \lesssim 3.0 \right) \; .
\label{eq:a3}
\end{equation}
Notice that in the kinematic regime where 3N
correlations are expected to dominate $\left( 2.25 \lesssim x_{B} \right)$
the experimental situation is unsettled.  For example, the recently
measured $a_{3}(^{4}$He/$^{3}$He) ratios \cite{PhysRevLett.108.092502} are
significantly larger than those reported in
Ref.~\cite{PhysRevLett.96.082501}.

\begin{figure}
 \includegraphics[width=\columnwidth]{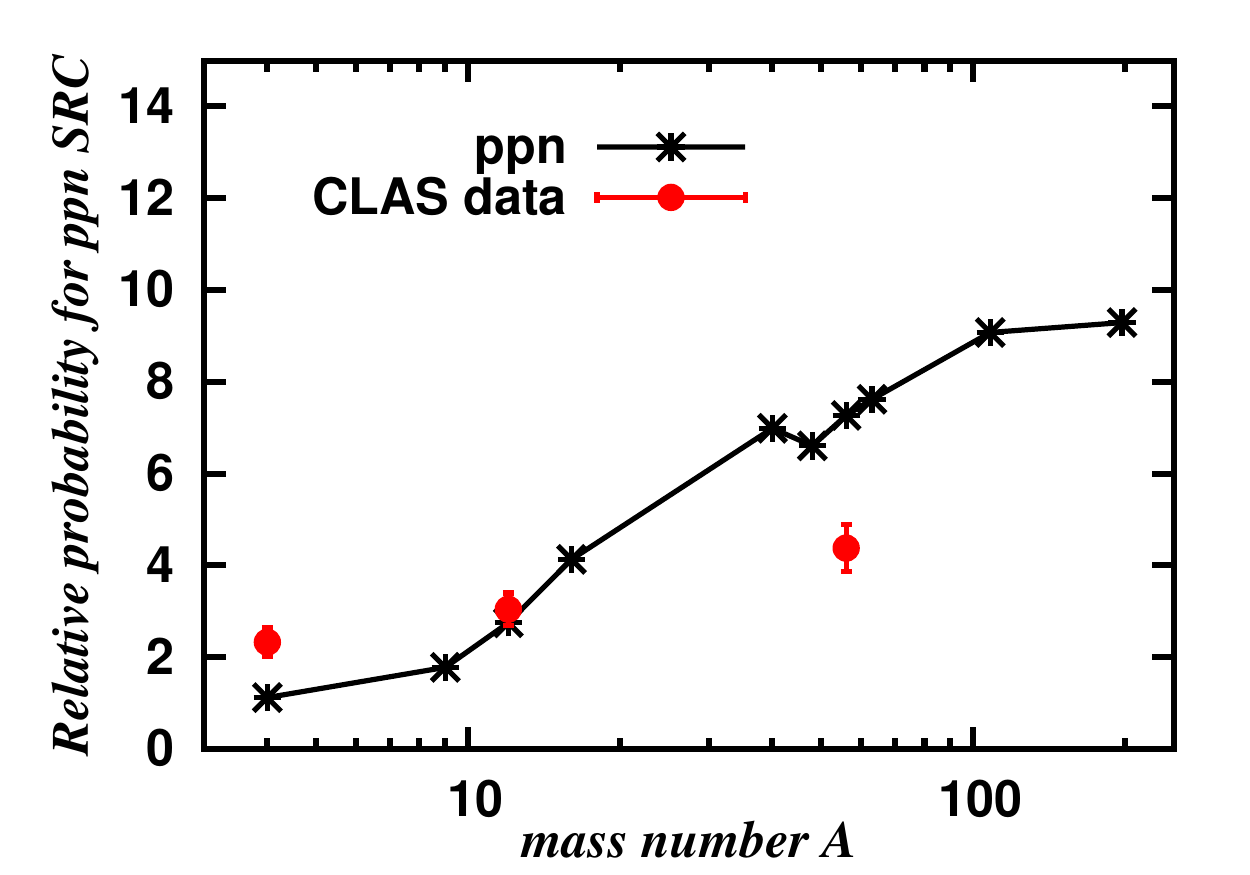}
 \caption{The mass dependence of the per nucleon probability for ppn
   SRC relative to $^{3}$He. We stress that neither the data nor the
   theoretical calculations have been corrected for c.m. motion and
   FSI effects. The data are from Ref.~\cite{PhysRevLett.96.082501}.}
\label{fig:relto3he}
\end{figure}

Similar to the per nucleon pn SRC of
Eq. (\ref{eq:pernucleonSRCrelativetod}) we define the per nucleon
probability for a ppn SRC relative to $^3$He as
\begin{equation}
\frac{3}{A} \frac{ N_{ppn} \left( A, Z \right) }
{ N_{ppn} \left( A=3 , Z=2 \right) } = 
\frac{3}{A} N_{ppn} \left( A , Z \right),
\label{eq:pernucleonppnSRC}
\end{equation}
where we used the fact that $N_{ppn} \left( A=3 , Z=2 \right) = 1$ in
our framework. The results of the per nucleon probability of ppn SRC are
collected in Figure~\ref{fig:relto3he}.

The quantity of Eq.~(\ref{eq:pernucleonppnSRC}) can be linked to
$a_3(A/^3\textrm{He})$ under the condition that corrections stemming
from c.m. motion of the correlated ppn triples, FSI effects, $\ldots$
are small. Under those idealized conditions one would have
\begin{equation}
a_3(A,^3\textrm{He}) \approx \frac {3} {A} N_{ppn} (A,Z) \; .
\label{eq:oura3definition}
\end{equation}
In the naive assumption that all 3N pairs contribute to the
$a_3(A/^3\textrm{He})$ ratio, one expects an $A^2$ dependency.  We
suggest that only ppn triples in a ``close'' configuration contribute
and we count the number of SRC triples with the aid of the
Eq.~(\ref{eq:project3NSRC}).  The ppn contributions will be larger
than the pnn ones due to the magnitude of the electromagnetic
coupling. Correlated triples should have at least one pn pair due to
the dominant character of the tensor component.  In
Fig.~\ref{fig:relto3he} we show the predictions for the
$a_3(A/{^3\textrm{He}})$ coefficient as computed with the Eq.~
(\ref{eq:oura3definition}) and compare it to the data.  We stress that
the experimental situation is largely unsettled and that neither the
data nor the theoretical calculations have been corrected for
c.m. motion and FSI effects. For Helium and Carbon our predictions are
in line with the experimental value. For Iron the prediction is about
a factor of two larger than the experimentally determined ratio of
cross sections. Our parameter-free calculations reproduce the fact
that the mass dependence is much softer than the $A^{2}$ dependence
that one would expect on naive grounds.

%
\section{Conclusion}

We have provided arguments that the mass dependence of the magnitude
of the NN and NNN correlations can be captured by some approximate
principles.  Our method is based on the assumption that correlation
operators generate the correlated part of the nuclear wave function
from that part of the mean-field wave function where two nucleons are
``sufficiently close''.  This translates to computing those parts of
the two-nucleon and three-nucleon wave functions with zero relative
orbital momentum in order to identify short-range
correlated pairs and triples.

We have calculated the number of pn, pp and nn $l_{12}=0$ SRC pairs
and studied their mass and isospin dependence.  The $A$ dependence of
the magnitude of the pp, nn, and pn SRC manifests itself in a
power-law dependence.  We found a significant higher per nucleon SRC
probability for pn pairs than for pp and for nn.  To connect the
computed number of SRC pairs to the measured $a_2 \left( A /
  \textrm{D} \right) $ corrections are in order. Published
experimental data include the radiation and Coulomb corrections.  The
correction factor stemming from final-state interactions and from the
c.m. motion of the correlated pair, however, is far from established.
We proposed a method to estimate the c.m. correction factor based on
general properties of nucleon momentum distributions. Using Monte
Carlo simulation we find a correction factor of about $1.7 \pm 0.3$.
Our model calculations for $a_2$ are of the right order of magnitude
and capture the $A$-dependence qualitatively. For small $A$ our
predictions underestimate the data, while we do not find the same
degree of saturation for high $A$ that the (scarce) data seem to
suggest.

To compute the number of 3N SRC in a nucleus, we count the ppn states
with three nucleons which are close.  We have quantified the number of
3N SRC and provided predictions for the measured $a_3 \left(
  A/^3\textrm{He} \right)$ coefficients.  Our model calculations for
the $a_3$ are of the same order of magnitude as the (scarce) data but
overestimate the $^{56}$Fe data point.  In this comparison no
corrections for c.m. motion and FSI effects have been made and it
remains to be studied in how far they can blur the connection between
inclusive electron-scattering data and the SRC information. We find a
linear relationship between the magnitude of the EMC effect and the
computed per nucleon number of SRC pn pairs. This may indicate that
the EMC effect is (partly) driven by local nuclear dynamics
(fluctuations in the nuclear densities), and that the number of pn SRC
pairs serves as a measure for the magnitude of this effect.

\subsection*{Acknowledgments}
The computational resources (Stevin Supercomputer Infrastructure) and
services used in this work were provided by Ghent University, the
Hercules Foundation and the Flemish Government – department EWI.  This
work is supported by the Research Foundation Flanders.

\bibliography{QuantifyingSRC.bbl}
\end{document}